\RequirePackage{fix-cm}
\documentclass[smallextended]{svjour3}       
\smartqed  
\usepackage{graphicx}
\journalname{Empirical Software Engineering}

\usepackage{natbib}
\usepackage{multirow}
\usepackage{subcaption}
\usepackage{makecell}
\usepackage{tcolorbox} 
\usepackage{eso-pic}
\usepackage{xcolor}
\usepackage[table]{xcolor}
\usepackage{comment}
\usepackage{listings}
\usepackage{hyperref}
\usepackage{booktabs}
\usepackage{amsmath}
\usepackage{amssymb}
\usepackage{breakcites}
\usepackage{url}

\usepackage{glossaries}

\newacronym{os}{OS}{Operating System}
\newacronym{nyi}{NYI}{Not Yet Implemented}
\newacronym{dei}{DEI}{Department of Informatics Engineering}
\newacronym{AI}{AI}{Artificial Intelligence}
\newacronym{BLEU}{BLEU}{Bilingual Evaluation Understudy}
\newacronym{ROUGE}{ROUGE}{Recall-Oriented Understudy for Gisting Evaluation}
\newacronym{METEOR}{METEOR}{Metric for Evaluation of Translation with Explicit ORdering}
\newacronym{ML}{ML}{Machine Learning}
\newacronym{NLP}{NLP}{Natural Language Processing}
\newacronym{ICL}{ICL}{In-Context Learning}
\newacronym{RAG}{RAG}{Retrieval Augmented Generation}
\newacronym{LSTM}{LSTM}{Long Short-Term Memory}
\newacronym{RNNs}{RNNs}{Recurrent Neural Networks}
\newacronym{BERT}{BERT}{Bidirectional Encoder Representations from Transformers}
\newacronym{MLM}{MLM}{Masked Language Modeling}
\newacronym{NSP}{NSP}{Next Sentence Prediction}
\newacronym{ASTs}{ASTs}{Abstract Syntax Trees}
\newacronym{MBPP}{MBPP}{Mostly Basic Python Problems}
\newacronym{RLHF}{RLHF}{Reinforcement Learning from Human Feedback}
\newacronym{GenAI}{GenAI}{Generative Artificial Intelligence}
\newacronym{CoT}{CoT}{Chain-of-Thought}
\newacronym{LLOC}{LLOC}{Logical Lines of Code}
\newacronym{SLOC}{SLOC}{Source Lines of Code}
\newacronym{LOC}{LOC}{Lines of Code}
\newacronym{MoE}{MoE}{Mixture of Experts}
\newacronym{GMM}{GMM}{Gaussian Mixture Models}
\newacronym{NLOC}{NLOC}{Number of Lines of Code}

\newglossaryentry{LLM}
{
  name={LLM},
  description={Large Language Model},
  first={Large Language Model (LLM)},
  plural={LLMs},
  descriptionplural={Large Language Models},
  firstplural={Large Language Models (LLMs)}
}

\usepackage{csquotes}

\newif\ifshowcomments
\showcommentstrue

\lstdefinestyle{smalljava}{
    language=Java,
    basicstyle=\ttfamily\fontsize{6pt}{6pt}\selectfont,
    backgroundcolor=\color{gray!10},
    frame=single,
    rulecolor=\color{gray},
    breaklines=true,
    showstringspaces=false,
    tabsize=2,
    captionpos=b,
    numbers=left,
    numberstyle=\tiny\color{gray},
    numbersep=5pt,
    xleftmargin=5pt,  
    xrightmargin=5pt
}

\lstdefinestyle{smallrust}{
    language=C++,
    basicstyle=\ttfamily\fontsize{6pt}{6pt}\selectfont,
    backgroundcolor=\color{gray!10},
    frame=single,
    rulecolor=\color{gray},
    breaklines=true,
    showstringspaces=false,
    tabsize=2,
    captionpos=b,
    numbers=left,
    numberstyle=\tiny\color{gray},
    numbersep=5pt,
    xleftmargin=5pt,  
    xrightmargin=5pt
}

\lstdefinestyle{smallcpp}{
    language=C++,
    basicstyle=\ttfamily\fontsize{6pt}{6pt}\selectfont,
    backgroundcolor=\color{gray!10},
    frame=single,
    rulecolor=\color{gray},
    breaklines=true,
    showstringspaces=false,
    tabsize=2,
    captionpos=b,
    numbers=left,
    numberstyle=\tiny\color{gray},
    numbersep=5pt,
    xleftmargin=5pt,  
    xrightmargin=5pt
}

\lstdefinestyle{smallpy}{
    language=Python,
    basicstyle=\ttfamily\fontsize{6pt}{6pt}\selectfont,
    backgroundcolor=\color{gray!10},
    frame=single,
    rulecolor=\color{gray},
    breaklines=true,
    showstringspaces=false,
    tabsize=2,
    captionpos=b,
    numbers=left,
    numberstyle=\tiny\color{gray},
    numbersep=5pt,
    xleftmargin=5pt,  
    xrightmargin=5pt
}

\lstdefinestyle{smallprompt}{
    basicstyle=\ttfamily\fontsize{6pt}{6pt}\selectfont,
    backgroundcolor=\color{gray!10},
    frame=single,
    rulecolor=\color{gray},
    breaklines=true,
    breakindent=0pt,
    showstringspaces=false,
    captionpos=b,
    numberstyle=\tiny\color{gray},
    numbersep=5pt,
    xleftmargin=5pt,
    xrightmargin=5pt,
    keepspaces=true,
    columns=fullflexible
}

\AddToShipoutPictureFG*{%
  \AtPageUpperLeft{%
    \raisebox{-1.0cm}{%
      \makebox[\paperwidth][c]{%
        \parbox{0.9\paperwidth}{%
          \centering
          \small\itshape
          Preprint version of a manuscript accepted for publication in
          \textit{Empirical Software Engineering}  (Springer).
        }%
      }%
    }%
  }%
}

\begin{document}

\title{PROBE: Benchmarking Code Generation in Large Language Models
\thanks{This work was partially supported by \enquote{Projeto n.º 2024.07660.IACDC, https://doi.org/10.54499/2024.07660.IACDC, apoiado pela medida “RE-C05-i08.M04 – “Apoiar o lançamento de um programa de projetos de I\&D orientado para o desenvolvimento e implementação de sistemas avançados de cibersegurança, inteligência artificial e ciência de dados na administração pública, bem como de um programa de capacitação científica”, do Plano de Recuperação e Resiliência – PRR, enquadrado no contrato de financiamento celebrado entre a Estrutura de Missão Recuperar Portugal (EMRP) e a Fundação para a Ciência e a Tecnologia I.P. (FCT), enquanto beneficiário intermediário.}}
}


\author{Rodrigo Pato Nogueira \and Marco Vieira \and João R. Campos}


\institute{Rodrigo Pato Nogueira
            \at \textit{University of North Carolina at Charlotte}, Charlotte, NC, USA
            \at \textit{University of Coimbra}, CISUC/LASI, DEI, Coimbra, Portugal \\
            \email{rpatodec@charlotte.edu}
           \and
           Marco Vieira \at
            \textit{University of North Carolina at Charlotte}, Charlotte, NC, USA \\
            \email{marco.vieira@charlotte.edu}
           \and
           João R. Campos \\
           \textit{University of Coimbra}, CISUC/LASI, DEI, Coimbra, Portugal \\
           \email{jrcampos@dei.uc.pt}
}

\date{Received: date / Accepted: date}

\maketitle

\begin{abstract}
\glspl{LLM} are increasingly being used in everyday software engineering tasks, particularly in automated code generation. Despite their widespread adoption, these models remain far from perfect, making systematic and fair evaluation essential to understand their strengths and limitations. In the context of code generation, existing benchmarks are limited: they often target a single programming language and rely primarily on unit test outcomes, while overlooking other critical dimensions such as the overall quality of the generated code and its closeness to a valid solution. To address these gaps, we introduce \texttt{PROBE}, an extensible benchmark framework that, unlike prior work, establishes a systematic structure built on diverse and well-defined metrics, representative workloads, varied prompt templates, and a robust experimental procedure. In practice, the code generated by the \glspl{LLM} is evaluated along three complementary dimensions: functional correctness, proximity to valid solutions, and code quality, enabling a comprehensive assessment of performance. We use \texttt{PROBE} to evaluate four open-source and two proprietary models under three prompting strategies across five programming languages. We further complement this analysis with a study of common errors in the code and provide concrete examples, offering clearer insight into where \glspl{LLM} tend to struggle. Our findings show that, while \glspl{LLM} achieve promising results, they struggle with harder problems and, in the case of smaller models, with programming languages that have fewer available resources for training, and they often fail due to fundamental and easily avoidable errors that underscore the unreliability of automatically generated code.
\keywords{Large Language Models, Benchmarking, Code Generation, Program Synthesis}
\end{abstract}

\section{Introduction}
\label{sec:intro}

The past few years have seen rapid progress in \glspl{LLM}, enabled by advances in model scale, training data, and architectural innovations \citep{chen2021evaluatinglargelanguagemodels,guo2025deepseek,openai2024gpt4technicalreport}. These models now demonstrate strong multitask capabilities and are widely applied in areas such as writing assistance~\citep{chen2025LLMWrintingAssistants}, summarization~\citep{fang2025multillmtextsummarization}, question answering~\citep{zhuang2023LLMsQA}, and software development~\citep{aleithan2024swebenchenhancedcodingbenchmark}.

A particularly impactful application is the generation of code from natural language descriptions, a task known as text-to-code generation~\citep{lu2021codexgluebenchmark}. This capability, exemplified by systems such as GitHub Copilot~\citep{github_copilot} and ChatGPT~\citep{openai_chatgpt}, is increasingly shaping developer workflows, with studies pointing to clear productivity gains in practical settings~\citep{ziegler2022copilotproductivity}. Still, the reliable generation of complete, correct, and verifiable software remains an open challenge. Recent evaluations show that \glspl{LLM} frequently underperform on complex or open-ended tasks~\citep{hendrycks2021APPS,khan2023xcodeeval} and are widely regarded as not yet \enquote{ready-to-use} for code generation in software engineering~\citep{zheng2025towards}.
These limitations highlight the need for benchmarks that systematically evaluate \glspl{LLM} on code generation, providing a clear picture of both their strengths and weaknesses.

Existing benchmarks for text-to-code generation~\citep{lu2021codexgluebenchmark,chen2021evaluatinglargelanguagemodels,austin2021programsynthesislargelanguage,khan2023xcodeeval,hendrycks2021APPS} have advanced the field but still present key limitations. Many target only a single programming language~\citep{chen2021evaluatinglargelanguagemodels,austin2021programsynthesislargelanguage,lu2021codexgluebenchmark} and, more generally, evaluations are often restricted to checking whether the generated code passes a predefined set of unit tests~\citep{chen2021evaluatinglargelanguagemodels,hendrycks2021APPS,khan2023xcodeeval,austin2021programsynthesislargelanguage}, while neglecting other crucial dimensions such as the proximity of the output to a correct solution and the overall quality of the produced code. Furthermore, these benchmarks often fail to adhere to established benchmarking practices, typically relying on limited datasets and lacking a clear definition of other essential components, such as a well-specified evaluation procedure~\citep{chen2021evaluatinglargelanguagemodels,hendrycks2021APPS,khan2023xcodeeval,austin2021programsynthesislargelanguage,lu2021codexgluebenchmark}. This hinders reproducibility and constrains the potential for extending the benchmarks to encompass additional aspects of code generation.

In this paper, we propose \texttt{PROBE}, a benchmarking framework that offers a systematic and extensible methodology for evaluating \glspl{LLM} on text-to-code generation tasks. \texttt{PROBE} integrates a rich set of evaluation metrics that go beyond unit test outcomes by also measuring the proximity of generated code to valid reference solutions and assessing its overall quality. Moreover, it incorporates a diverse workload of programming problems drawn from the IBM CodeNet dataset~\citep{puri2021codenet}, systematic prompt templates, and a well-defined evaluation procedure. These components provide a foundation for reproducible, fair, transparent, and comprehensive assessment of model capabilities.

We applied \texttt{PROBE} to a set of representative (and widely used) open-source models (Qwen2.5:14b, Qwen2.5-Coder:14b, Qwen2.5-Coder:7b, and Deepseek-Coder-v2:16b) and proprietary \glspl{LLM} (GPT-4.1-mini and Gemini-2.0-flash) across five programming languages: Python, C++, Java, C, and Rust. These languages were selected for their contrasting syntactic characteristics and varied representation in training data. Evaluations were conducted under three prompting strategies: baseline (no additional context), \gls{ICL} (1-shot learning), and feedback-incorporation (two rounds of iterative refinement). Our results show that model performance varies substantially across programming languages. Models generally perform better in Python, C++, and Java than in C and Rust, with smaller open-source models struggling particularly in low-resource languages such as Rust. Feedback-incorporation consistently improved the functional correctness of the code, whereas \gls{ICL} yielded only marginal gains. Notably, just as functional correctness and proximity to valid solutions drastically decline with increasing problem difficulty, the same does not happen with the gap in code quality between LLM-generated and human-written code. These findings reveal trade-offs in generated code and underscore the importance of robust benchmarking frameworks such as \texttt{PROBE}.

To complement these quantitative results, we further perform an analysis of the generated code to identify common failure patterns and provide illustrative examples of where models tend to struggle. These results showed that the generated code is often unreliable, exhibiting fundamental and easily avoidable errors such as missing imports, integer overflows, accessing arrays out of bounds, and relying on brute-force algorithms that do not scale. These issues highlight the risk of overtrusting \glspl{LLM} for code generation, as such flaws can lead to vulnerabilities or performance regressions in real systems. 

In summary, the main contributions of this paper are:

\begin{itemize}

    \item \textbf{\texttt{PROBE}}, a novel, systematic, and extensible framework for benchmarking \glspl{LLM} in text-to-code generation. Unlike prior efforts that rely on a narrow set of metrics or ad hoc setups, \texttt{PROBE} delivers a systematic methodology that evaluates models across three different dimensions, ensuring consistency, extensibility, and reproducibility across programming languages, models, and prompting strategies.

    \item A \textbf{large-scale evaluation} covering six \glspl{LLM}, five languages with distinct syntactic and semantic challenges, and three prompting paradigms. This analysis reveals systematic variations in performance across models, languages, and prompting strategies, offering a clearer understanding of current strengths and limitations. 

    \item \textbf{Extensibility guidelines} detailing how to incorporate new workloads, metrics, and prompting methods, ensuring the framework is extensible and adaptable to future text-to-code generation demands.

    \item A detailed \textbf{Error Analysis}, including an overview of the most common error causes along with illustrative examples, to provide a clearer understanding of where the \glspl{LLM} tend to struggle.
    
    \item PROBE project page\footnote{\url{https://huggingface.co/datasets/OSS-forge/PROBE}}, a \textbf {public repository} where the workload and evaluation results are continuously released. This resource enhances transparency, enables fair model comparisons, and lowers the barrier to extending the framework with new workloads or evaluation dimensions. 
    
\end{itemize}

This paper is structured as follows. Section~\ref{sec:background-relatedwork} reviews background and related work. Section~\ref{sec:framework} presents the proposed framework, Section~\ref{sec:experimental} the experimental setup, and Section~\ref{sec:results} the results across models and programming languages. Section~\ref{sec:dicusiion-ttv} discusses the results and the threats to validity, and Section~\ref{sec:conclusion} concludes with directions for future research.

\section{Background and Related Work}
\label{sec:background-relatedwork}

\glspl{LLM} are increasingly reshaping software engineering by supporting tasks such as automated code generation, bug identification, and code translation~\citep{lu2021codexgluebenchmark}. These models, trained on large amounts of data, are capable of learning patterns that enable them to tackle complex programming challenges, such as generating code from natural language descriptions. However, their performance remains limited, with even the biggest models struggling with more complex problems~\citep{hendrycks2021APPS, lu2021codexgluebenchmark}. Evaluating the quality of code generated by \glspl{LLM} is therefore critical. Beyond checking whether the output is functionally correct, a comprehensive evaluation should also measure how close the generated code is to a correct solution and the overall quality of the implementation.

Functional correctness is typically measured through execution-based metrics. Compilation or interpretation success rate, for instance, reflects whether the generated code can be executed without syntax errors~\citep{wang2022compilableneuralcodegeneration}. On the other hand, \textit{Test Case Average} reflects the average fraction of passed unit tests across problems~\citep{hendrycks2021APPS}. The most widely adopted metric, however, is \textit{pass@k~}\citep{chen2021evaluatinglargelanguagemodels, austin2021programsynthesislargelanguage, khan2023xcodeeval}, which evaluates multiple (k) generated samples per problem and estimates the probability that at least one is fully correct, meaning it passes all unit tests.

While execution-based metrics capture correctness, they do not measure how close incorrect attempts are to a valid solution. For this purpose, similarity-based metrics are employed. Traditional metrics such as BLEU \citep{BLEU_papinesi} and ROUGE \citep{lin2004rouge} have been used, but the most common in code evaluation is CodeBLEU \citep{ren2020codebleu}. This metric combines n-gram matching with syntactic and semantic similarity, as well as dataflow-based comparisons, offering a comprehensive view of solution proximity.

A further dimension of evaluation concerns the quality of the generated code itself. This can be assessed through static measures that do not require execution, such as cyclomatic complexity, which captures the number of independent paths in a program’s control flow graph, or \gls{NLOC}, which counts the number of executable lines of code and provides an estimate of program size. These metrics highlight aspects of conciseness and maintainability, but they provide only indirect indications of overall code quality. More detailed insights can be obtained through feedback-based metrics, where humans evaluate properties such as readability, maintainability, and practical applicability~\citep{chen2024surveyevaluatinglargelanguage}. Although these assessments capture nuances beyond what static analysis can measure, they are subjective and require considerable human effort.

Benchmarking is a standardized approach to evaluate and compare systems under consistent conditions, guided by well-defined workloads, metrics, and procedures~\citep{gray1992benchmark}. At its core, it constitutes an experimental methodology that enables reproducible measurement of key properties such as performance, dependability, and security~\citep{gray1992benchmark, vieira2003dependability}. The effectiveness and acceptance of a benchmark ultimately depend on how accurately it reflects real-world conditions and its ability to produce consistent and comparable results. To ensure a meaningful and trustworthy evaluation of \glspl{LLM} for text-to-code generation, a benchmark must satisfy several essential properties~\citep{gray1992benchmark, vieira2003dependability}. These include \textbf{ease of installation and use}, ensuring the benchmark is straightforward to deploy with minimal setup effort; \textbf{promptness}, as efficient execution reduces costs and improves usability; and \textbf{non-intrusiveness}, which guarantees the benchmark does not alter the internal behavior of the models being tested, preserving result validity. \textbf{Portability} is also critical, enabling the benchmark to run across different platforms without significant modifications and ensuring fair comparisons. Equally important are \textbf{repeatability}, which ensures consistent results across multiple executions under the same conditions, and \textbf{representativeness}, which guarantees results reflect realistic usage scenarios. These properties ensure a benchmark is not only rigorous and practical but also relevant for guiding the selection and deployment of models in real-world settings.

In recent years, multiple benchmarks for \glspl{LLM} on text-to-code generation have been proposed~\citep{khan2023xcodeeval, lu2021codexgluebenchmark, chen2021evaluatinglargelanguagemodels, austin2021programsynthesislargelanguage, hendrycks2021APPS}. While these efforts have helped advance the field, important gaps remain. Many benchmarks focus on a single programming language (most often Python)~\citep{hendrycks2021APPS, austin2021programsynthesislargelanguage, chen2021evaluatinglargelanguagemodels}. Although this choice simplifies evaluation, it overlooks language-specific failure modes and restricts insights into model robustness in realistic multilingual scenarios. Moreover, most benchmarks report performance exclusively through the pass@k metric~\citep{chen2021evaluatinglargelanguagemodels, khan2023xcodeeval, hendrycks2021APPS}. While this metric captures whether generated code is entirely correct, it does not indicate partial correctness, the proximity of an incorrect solution to a valid one, or the overall quality of the code. Finally, many existing benchmarks are little more than datasets~\citep{khan2023xcodeeval, chen2021evaluatinglargelanguagemodels, hendrycks2021APPS}, lacking a clearly defined evaluation procedure and failing to demonstrate adherence to fundamental benchmarking principles.

The \texttt{PROBE} framework addresses these limitations by building on a dataset of real-world problems spanning multiple difficulty levels. It is language-agnostic and evaluates code along complementary dimensions, thereby offering a more comprehensive view of model performance. Furthermore, its structured, scalable, and extensible design enables consistent and fair assessment across both models and programming languages.

\section{The \texttt{PROBE} framework}
\label{sec:framework}

\texttt{PROBE} is a structured benchmarking framework for the systematic evaluation of \glspl{LLM} on text-to-code generation tasks. Its design is grounded in core benchmarking principles, aiming to ensure representativeness through real-world problems, guarantee repeatable results, and remain non-intrusive, preserving natural model behavior. The framework supports workload scalability, provides partial portability across models and tasks (while acknowledging that full portability, particularly in terms of installability and replaceability, is outside the scope), and promotes simplicity through automation. \texttt{PROBE} has a clear evaluation scope and is built with extensibility in mind, ensuring continued relevance as models, metrics, workflows, prompts, and deployment settings evolve.

\begin{figure}
    \centering
    \includegraphics[width=1.0\linewidth]{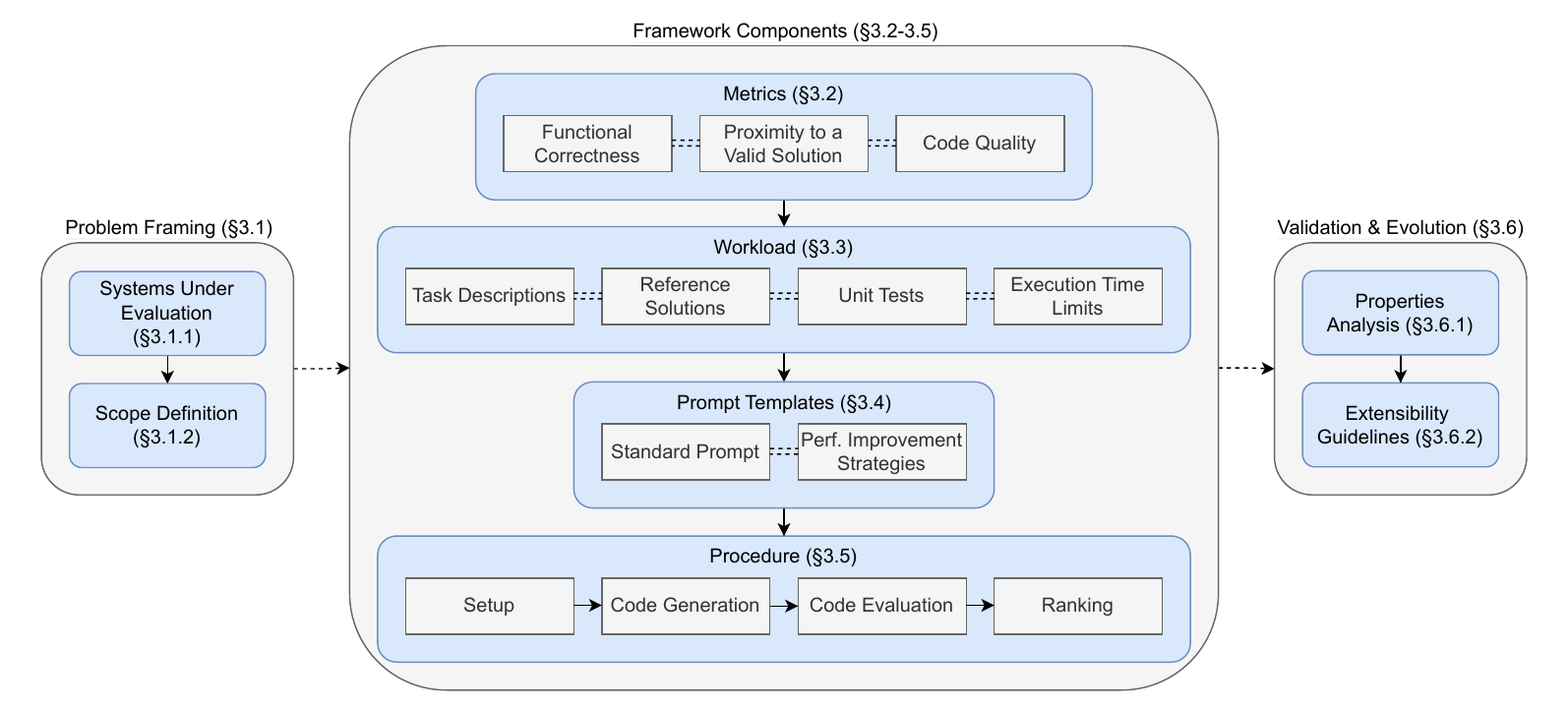}
    \caption{Design of the PROBE Benchmarking Framework.}
    \label{fig:framework}
\end{figure}

The framework is structured around three core elements. As shown in Figure~\ref{fig:framework}, the first is \textbf{Problem Framing}, which establishes the foundation for the benchmark design by clearly defining the systems under study and clarifying the scope of the framework. The second is \textbf{Framework Components}, which encompass the essential components required to assess \glspl{LLM} accurately:

\begin{itemize}

    \item \textit{Metrics}: specify how model performance is evaluated, including functional correctness, proximity to a valid solution, and overall code quality.

    \item \textit{Workload}: defines the resources required to generate and evaluate code, including problem descriptions, reference solutions, unit tests, and execution time constraints.

    \item \textit{Prompt Templates}: templates used to query the models during evaluation, including both standard prompts and variations designed to explore performance-improvement strategies.

    \item \textit{Procedure}: defines the steps of the benchmarking process, namely prompt formatting, model execution, evaluation, and model ranking.

\end{itemize}

The third element, \textbf{Validation and Evolution}, serves two main purposes. It includes an analysis of the key benchmarking properties to ensure the evaluation remains robust and meaningful across different scenarios, and provides practical guidelines for extending the framework to incorporate new languages, metrics, workloads, or prompts.

Although our approach is intrinsically empirical, it is grounded in well-established benchmarking practices. As with any evaluation framework, \texttt{PROBE} abstracts practical use cases into a controlled setting, and direct generalization to all real-world scenarios is neither intended nor supported. For example, our evaluation considers a specific set of programming languages and workloads, so conclusions should be interpreted within the scope of these languages, tasks, and problem sets.

\subsection{Problem Framing}
\label{sec:problemframing}

Problem framing defines the context in which the evaluation takes place. This involves characterizing the systems under evaluation and specifying the scope of the assessment.

\subsubsection{Systems Under Evaluation}
\label{sec:sue}

Characterizing the systems under evaluation is essential. For the task of code generation, this means specifying how \glspl{LLM} are expected to produce code from natural language descriptions and what assumptions guide their use. Such characterization ensures that the framework reflects realistic application settings while enabling structured and consistent comparisons across models. In our case, we assume a black-box interaction model, where systems are accessed exclusively through prompts, without modifying their internal components. This reflects how \glspl{LLM} are commonly deployed in practice and motivates a non-intrusive evaluation.

\subsubsection{Scope Definition}
\label{sec:scope}

Clearly defining the evaluation scope is key for ensuring a structured and meaningful assessment. This includes specifying the task type and the performance-improvement strategies to be applied (e.g., \gls{ICL}). Establishing scope not only guarantees consistency, reproducibility, and interpretability across tasks and models but also determines the artifacts required to construct an appropriate workload. For instance, solving a problem from scratch may only need a self-contained description. In contrast, tasks involving integration with an existing codebase may require additional context, e.g., function signatures, constraints, or surrounding code.

\texttt{PROBE} focuses on the task of generating complete programming problems, where models are required to create standalone programs that take inputs and produce outputs. This evaluation setting compels models to capture and integrate multiple aspects of the problem specification, including input/output behavior, constraints, and overall program structure. Such problems provide a realistic and rigorous testbed, as they closely mirror real-world programming scenarios more than isolated code snippets, and therefore offer more insight into model capabilities and limitations.

For performance-improvement strategies, \texttt{PROBE} considers two approaches in addition to a no-strategy baseline. The first is \gls{ICL}, where an input–output example is included in the prompt to better guide the model. The second is feedback incorporation, where the generated code is executed and, if incorrect, the model receives error feedback and may revise its output up to the number of iterations specified by the benchmark user. These strategies represent widely used techniques for enhancing LLM performance: \gls{ICL} provides models with minimal task demonstrations, while feedback-driven refinement captures interactive scenarios that more closely reflect practical development workflows.

If needed, the scope of this benchmark can be easily extended to cover broader tasks (e.g., multi-step problems, domain-specific workloads) by following the extensibility guidelines outlined in Section~\ref{sec:extensibility}.

\subsection{Metrics}
\label{sec:metrics}

Selecting appropriate metrics is critical in any benchmarking procedure, as they define performance measurement and enable fair model comparisons. When evaluating how well an \gls{LLM} generates code from a natural language description, the main goal is to determine whether the code performs the intended task. However, correctness alone is not sufficient: a solution might fail some tests yet demonstrate understanding or follow a valid structure, while a correct solution could be hard to read or maintain. Therefore, a comprehensive evaluation should consider \textbf{functional correctness}, \textbf{proximity to a valid solution}, and \textbf{quality of the generated code}, including both behavior and structure. Below, we present the metrics for each dimension.

\subsubsection{Functional Correctness}
\label{sec:metrics-func}

The evaluation of functional correctness is based on the execution results of the generated code. Metrics must capture not only complete correctness but also partial correctness, enabling a more fine-grained analysis of model performance. To this end, the \texttt{PROBE} benchmark uses two execution-based metrics: \textit{pass@k} and \textit{Outcome Rate}.

The \textit{pass@k} metric estimates the probability that at least one of the top-$k$ generated solutions is fully correct. It can be computed using its unbiased estimator~\citep{chen2021evaluatinglargelanguagemodels}:

\begin{equation}
    \text{pass@}k := \mathbb{E}_{\text{Problems}} \left[ 1 - \binom{N-c}{k}/\binom{N}{k} \right]
\end{equation}

\noindent where N is the number of samples generated for each problem and c is the number of samples that successfully pass all unit tests. 

The \textit{Outcome Rate} classifies each generated solution into one of six outcomes (Passed, Failed, Timeout, Error, No Compile, and No Code) and measures the average proportion of executions associated with each outcome. It is computed as follows:

\begin{equation}
    \text{OutcomeRate}_{o} := \frac{1}{P} \sum_{p=1}^{P} \left( \frac{1}{N} \sum_{n=1}^{N} \left( \frac{t_{p,n}^{(o)}}{T_p} \right) \right)
\end{equation}

\noindent where $P$ is the total number of problems, $N$ the samples per problem, $T_p$ the number of unit tests for problem $p$, and $t_{p,n}^{(o)}$ the number of executions for generation $n$ of problem $p$ that result in outcome $o$. For instance, if a generated code sample passes one of four unit tests and fails the other two due to execution errors, the outcome rates would be $0.33$ for Passed and $0.66$ for Error. If the code does not compile or if no code is generated, the outcome rate is set to $1$ for No Compile/No Code and $0$ for all other outcomes. This metric provides a more detailed view of model behavior than binary correctness alone, allowing us to differentiate between partially and completely correct solutions.

\subsubsection{Proximity to a Valid Solution}
\label{sec:metrics-prox}

Similarity-based metrics are used to assess how close a generated solution is to a correct one. In the \texttt{PROBE} benchmark, we use the CodeBLEU score~\citep{ren2020codebleu}, as this metric combines multiple aspects of code similarity into a single measure, and prior empirical studies have shown it to correlate more strongly with human judgments than traditional metrics such as BLEU across several code generation tasks~\citep{ren2020codebleu}. This metric is defined as a weighted sum of four components: n-gram matching, weighted n-gram matching, syntactic AST match, and semantic dataflow match. Following prior work~\citep{ren2020codebleu}, the weights are set to 0.1 for both n-gram components and 0.4 for the syntactic and semantic components, as this configuration has been shown to yield the strongest correlation with human judgments in text-to-code generation tasks.

For each generated sample, the CodeBLEU score is computed against all available reference implementations, and the maximum value is retained to account for the existence of multiple valid solutions. To aggregate across samples and problems, the score is averaged over all generated samples for a problem and then across all problems:

\begin{equation}
\text{CodeBLEU} := \frac{1}{P} \sum_{i=1}^{P} \left( \frac{1}{N} \sum_{n=1}^{N} \max_{j \in \{1, \dots, M_i\}} \text{CodeBLEU}(\hat{C}_{i,n}, C_{ij}) \right)
\end{equation}

\noindent where $\{C_{ij}\}_{j=1}^{M_i}$ are the reference solutions for problem $i$ and $\hat{C}_i$ is the generated sample. 

We recognize that, by calculating CodeBLEU in this way, we may penalize generations that are perfectly correct but adopt an approach different from all reference solutions (e.g., using an alternative data structure or algorithmic strategy). To mitigate this, we ensure multiple reference solutions per problem, reducing the likelihood that a valid solution diverges from all references. An alternative approach, such as computing CodeBLEU only for incorrect generations, would avoid penalizing correct solutions but would introduce other issues: each model would be evaluated on a different subset of problems (those it failed), making normalization difficult. Models that miss longer or more complex problems would be disproportionately penalized because n-gram similarity is inherently sensitive to token count. 

Despite these limitations, we retain CodeBLEU as a complementary metric because it captures aspects of solution similarity that are not reflected in functional correctness or static code-quality measures. In our experiments (see section~\ref{sec:results-proximity}), we observed that CodeBLEU scores are broadly aligned with functional correctness at the model level, suggesting that, while CodeBLEU cannot replace execution-based evaluation and may penalize structurally different correct solutions, it still captures meaningful patterns in how models approach problems. It also provides insight into whether generated solutions follow common structural and semantic patterns in the reference implementations, enabling a more complete characterization of model behavior.

\subsubsection{Code Quality}
\label{sec:metrics-quality}

Code quality should be evaluated using metrics that provide a meaningful characterization of the generated solutions. In the \texttt{PROBE} benchmark, two static analysis metrics are considered: cyclomatic complexity and \gls{NLOC}. The first reflects the structural complexity and potential difficulty of understanding the code, while the second captures its length and conciseness.

To avoid misleading results, code quality metrics should be computed only for correct solutions, i.e., those that pass all unit tests. Evaluating incorrect code could be misleading, as simpler but wrong implementations may appear artificially clean. Furthermore, solutions are compared to the reference implementations in relative rather than absolute terms, ensuring that models tackling more complex problems are not unfairly penalized. For each problem $i$, let $\overline{M}_{ref, i}$ and $\overline{M}_{gen, i}$ denote the mean value of metric $M \in \{\text{cyclomatic complexity, nloc}\}$ for the reference solutions and for the correct generated samples, respectively. If a problem has no correctly generated samples, the metric is not computed. To account for both absolute and relative differences, we define:

\begin{equation}\label{eq:abs_score}
    score_{M, i} = 
    \begin{cases} 
        1 - e^{-\beta_i (\overline{M}_{gen, i} - \overline{M}_{ref, i})^2}, & \overline{M}_{gen, i} \leq \overline{M}_{ref, i} \\[4pt]
        e^{-\gamma_i (\overline{M}_{gen, i} - \overline{M}_{ref, i})^2} - 1, & \overline{M}_{gen, i} > \overline{M}_{ref, i} 
    \end{cases}
\end{equation}

\noindent where the sensitivity parameters (\(\beta_i\) and \(\gamma_i\)) are scaled based on the reference metric:

\begin{equation}
    \beta_i = \frac{1}{\overline{M}_{ref,i}}, \quad \gamma_i = \frac{1}{\overline{M}_{ref,i}}
\end{equation}

The overall metric $score_M$ is obtained by averaging $score_{M,i}$ over all problems with at least one correct solution. This scaling measures deviations relative to the reference value, capturing both absolute and proportional differences. In this way, the score reflects raw magnitude (e.g., 50 vs 60 versus 5 vs 6) while accounting for proportionality (e.g., 50 vs 51 is less significant than 5 vs 6).

Although many other metrics could be included (e.g., code smells, readability scores, or broader maintainability indices), we chose not to include them in \texttt{PROBE}. Static metrics of this kind are hard to apply consistently across programming languages because they often depend on language-specific conventions, linter rules, or static-analysis tools that can disagree even within the same ecosystem, making their outputs heavily tool-dependent~\citep{lenarduzzi2023critical}. Likewise, feedback-based evaluation and runtime performance were also not included, as these metrics cannot be automatically computed in a controlled and reproducible way.  Feedback-based metrics rely on human judgment, making them subjective and difficult to standardize, while execution time cannot be reliably measured without exclusive control over the runtime environment. In contrast, cyclomatic complexity and \gls{NLOC} provide comparatively stable, tool-agnostic measurements that can be computed reliably across languages, making them suitable for a uniform and fully automated benchmark. Although we do not include the additional metrics above, they can be integrated into future extensions of the framework (see Section~\ref{sec:extensibility}).

\subsection{Workload}
\label{sec:workload}

The workload should provide all the elements necessary to both define code generation tasks and evaluate the generated code. To enable comprehensive assessment, it should span a range of difficulty levels and be built around four key components: task descriptions, reference solutions, unit tests, and execution time limits.

The workload used for \texttt{PROBE} is derived from the Project CodeNet dataset \citep{puri2021codenet}, a large collection of 4,053 problems from two different online judge websites, AIZU~\footnote{\url{https://onlinejudge.u-aizu.ac.jp/introduction}} and AtCoder~\footnote{\url{https://atcoder.jp/}}, along with over 13 million solutions across 55 programming languages. While CodeNet provides diversity and scale, its raw form is not directly suitable for benchmarking \glspl{LLM}. Therefore, each component required additional processing, as discussed in the following paragraphs. To assist with these steps, we used the Qwen2.5 model~\citep{hui2024qwen25codertechnicalreport} (72B parameters, 4-bit quantized), which offers strong performance while remaining feasible to run on our available hardware.

We acknowledge a potential concern regarding the dataset's representativeness. However, although IBM CodeNet originates from competitive programming platforms, it includes problems spanning a wide variety of structures, complexity levels, and programming constructs. Many tasks involve I/O handling, memory management, and control flow logic, challenges that are central to text-to-code generation. While the workload differs from large-scale industrial projects in domain and scope, it provides a controlled and sufficiently diverse evaluation setting to reveal meaningful performance differences among \glspl{LLM}. Moreover, our results (see Section~\ref{sec:results}) demonstrate clear distinctions across models, prompt designs, and problem complexity. These findings suggest that even a limited yet well-structured benchmark can effectively expose both the strengths and limitations of current code generation capabilities.

\subsubsection{Task Descriptions}

Task descriptions specify the problem in detail, including input/output formats and constraints, ensuring that models have all the information required to solve the task. In \texttt{PROBE}, task descriptions were obtained from the original problem statements in the CodeNet dataset. However, two issues were identified:

\begin{itemize}

    \item \textbf{Language}: Some descriptions were written in Japanese, introducing potential language bias. To address this, each description was classified by the Qwen2.5 model as either Japanese or English. A total of 1,412 problems were identified as being in Japanese and were removed, rather than translated, to avoid introducing potential errors or semantic inconsistencies in the problem statements that could affect evaluation reliability.
    
    \item \textbf{Format}: All descriptions were stored in HTML, adding unnecessary complexity and potentially confusing models not trained to parse markup. To resolve this, task descriptions, input/output formats, and problem constraints were extracted using relevant HTML tags. To validate the extraction, the Qwen2.5 model was shown both the original and extracted versions and asked whether all relevant information was preserved and whether no input–output examples were mistakenly included. Any flagged cases were manually reviewed and corrected. During this process, 45 problems were found to lack sufficient information (e.g., referencing another problem without providing details). These problems were removed.
    
\end{itemize}

Although both steps rely on \glspl{LLM}, their role is limited to auxiliary preprocessing tasks, namely language classification and validation of extracted content. These operations do not modify the problem statements or introduce new information, and therefore do not affect the semantic content of the benchmark or the evaluation criteria.

After addressing both issues, we retained 2,596 problems, with descriptions organized in the format shown in Listing~\ref{lst:problem_desc}. This structured representation ensures that models can interpret tasks unambiguously and enables systematic evaluation.

\begin{figure}
\begin{lstlisting}[style=smallprompt, label={lst:problem_desc}, caption={Problem Description structure}]
Problem Description: {task description}
Input Format: {input format}
Output Format: {output format}
Constraints: {constraints}
\end{lstlisting}
\end{figure}

\subsubsection{Reference Solutions}\label{sec:workload-reference-sols}

Reference solutions are essential for the evaluation of the generated code along two of the dimensions: proximity to a valid solution and code quality. To ensure coverage of multiple representations, we recommend a minimum of three reference solutions per programming language for each problem (a greater number can obviously be used). If this minimum cannot be met, the metrics related to proximity to reference solutions and code quality must not be computed.

The reference solutions used in \texttt{PROBE} were obtained from the CodeNet dataset. However, once again, two problems were identified:

\begin{itemize}

    \item \textbf{Code Correctness}: The CodeNet dataset contains a large number of solutions, but many are incorrect. To address this, we applied a three-step validation procedure. First, we used the metadata provided in CodeNet and removed all solutions labeled as incorrect. Second, we leveraged the unit tests available in the dataset to perform an additional filtering step, discarding any solutions that failed these tests. Although these tests are limited and not sufficient for full evaluation, they help identify mislabeled submissions. Finally, after constructing a complete set of unit tests for each problem (see section~\ref{sec:workload-unit}), all remaining solutions were re-evaluated, and only those that passed every test were retained.

    \item \textbf{Code Quality}: Some solutions, while functionally correct, exhibited disproportionately high structural complexity compared to other implementations of the same problem. To identify them, rather than relying on subjective notions of code quality (e.g., readability or style), we use cyclomatic complexity as a reproducible proxy. For each problem, we compute the distribution of cyclomatic complexity across all solutions and exclude those whose complexity exceeds three standard deviations above the mean, a commonly used criterion for identifying statistical outliers. This filtering step targets only extreme cases (e.g., unnecessarily intricate control flow) that could bias downstream analyses, rather than typical variation in implementation style.

\end{itemize}

\subsubsection{Unit Tests}\label{sec:workload-unit}

Unit tests are used to validate the functional correctness of the code. To enable thorough performance evaluation, each problem must include at least four unit tests: three for assessing code correctness and one reserved as an example for \gls{ICL}. The smallest test is selected as the \gls{ICL} example to avoid exceeding the model's context window, since some tests can contain thousands of characters. Preliminary experiments on a subset of the dataset showed that varying the number (1--3) and selection of \gls{ICL} examples had no meaningful impact on overall performance, supporting the use of a single, consistent example. If more than four tests exist, one is reserved for \gls{ICL} and the rest for evaluation. Separating evaluation tests from \gls{ICL} examples prevents overfitting and inflated correctness metrics. Nevertheless, this design is not restrictive: following the extensibility guidelines described in Section~\ref{sec:extensibility}, both the number and selection of \gls{ICL} examples can be easily modified by adjusting the dataset and prompt templates.

Although CodeNet includes some unit tests, none of its problems meet the minimum requirement of four, with most providing only a single test. To complement the dataset, additional unit tests were collected from the original problem websites (AIZU and AtCoder). When such tests were unavailable, they were generated using the Qwen2.5 model, which was provided with the existing tests and input format and instructed to create new inputs following the same structure. Each generated input was validated against three independent reference solutions and retained only if all produced the same output; this process was repeated until four valid and consistent tests were obtained per problem.

Once this process was complete, excluding the unit tests reserved for \gls{ICL}, each problem in \texttt{PROBE} contained between 3 and 124 unit tests, with an average of 19.61 unit tests per problem.

Regarding input diversity, the LLM was prompted to generate inputs that covered different regions of the input space (compared to the already existing tests). However, we did not explicitly enforce edge-case generation. In some problems, relevant edge cases involve extremely large inputs that exceed the maximum output token limits of even large LLMs, making them infeasible to generate within our setup. We acknowledge that this may limit the ability of the test suites to fully capture all corner cases. To mitigate this limitation and ensure that the generated test suites remain sufficiently comprehensive, we conducted a systematic assessment of test adequacy using coverage metrics.

For each problem, we randomly sampled three reference implementations in Python, computed line, branch, and mutation coverage for each, and averaged the results across implementations and then across all problems. This approach reduces the impact of implementation-specific artifacts on the reported metrics. Overall, we obtained average values of 97.1\% line coverage, 94.1\% branch coverage, and 88.64\% mutation coverage, indicating that the generated test suites exercise a substantial portion of the program logic. A manual inspection of lower-coverage cases revealed that they are primarily due to implementation-specific factors (e.g., auxiliary or unused functions, defensive code paths, or rarely triggered branches), rather than deficiencies in the generated tests.

While generating synthetic unit tests carries some risk, this was done only for the subset of problems that lacked sufficient tests initially (255 out of 1,651). Although synthetic tests may influence coverage for this subset, we believe this is unlikely to meaningfully affect the overall trends observed, given the limited proportion of affected problems, the use of multiple reference implementations for validation, and the consistently high coverage observed across the dataset.

\subsubsection{Execution Time Limits}

Language-specific execution time limits must also be defined for each problem. These limits must also be adapted to the hardware used for evaluation, taking into account differences in language efficiency and hardware performance. Although the objective of this benchmark is not to evaluate whether \glspl{LLM} generate highly optimized or high-performance code, such limits remain necessary. Without them, the evaluation could stall due to infinite loops or extremely inefficient implementations. Performance-oriented metrics could be incorporated in the future if desired, following the framework’s extensibility guidelines (see Section~\ref{sec:extensibility}).

In \texttt{PROBE}, these limits were determined based on the performance of the reference solutions on the unit tests. Specifically, they were set to three times the duration of the slowest reference solution executing its longest-running unit test. While not overly restrictive, these limits effectively prevent the evaluation from stalling on infinite loops, while still providing enough time for models to generate correct solutions. Note that stricter limits may not be feasible, as complete control over the evaluation hardware may not be guaranteed.

\subsection{Prompt Templates}
\label{sec:prompts}

Prompt templates define the structure and content of inputs used to evaluate \glspl{LLM} in code generation tasks. They standardize interactions while allowing flexibility to explore different prompting strategies. Two main scenarios are considered: \textit{Standard Prompt}, where no performance-improving techniques are applied, and \textit{Performance Improvement Strategies}, which integrate specific prompting methods intended to enhance model performance. The prompt templates used are available at the PROBE project page\footnote{\url{https://huggingface.co/datasets/OSS-forge/PROBE}}.

\subsubsection{Standard Prompt}
\label{sec:prompts-standard}

Standard prompts serve as the baseline, with no performance-enhancing strategies applied. Each prompt must clearly state the task and target programming language, and include a detailed problem description. In the \texttt{PROBE} benchmark, the standard prompt is divided into two components: the system prompt and the user prompt.

The system prompt (Listing~\ref{lst:system_prompt}) contains the information that remains constant across all problems. It starts by assigning the role \enquote{You are an AI Code Assistant}, setting a helpful, code-focused tone to improve clarity and relevance. Next, a directive instructs the model to generate code following a specific structure, reducing ambiguity. Finally, the formatting requirement to wrap code within \texttt{[code]...[/code]} tags ensure consistency and simplify code extraction.
The user prompt (Listing~\ref{lst:user_prompt}) provides information specific to each problem. It includes the extracted problem description, input/output formats, and constraints, followed by a line specifying the target programming language. While this information also appears in the system prompt, early experiments showed that repeating it here helped reduce extraneous explanations and encouraged more direct outputs. 

\begin{minipage}{0.45\textwidth}
\begin{lstlisting}[style=smallprompt, caption={System prompt template}, label={lst:system_prompt}]
You are an AI Code Assistant.

You will be given a problem with structure:
Problems Description: ...
Input Format: ...
Output Format: ...
Constraints: ...

Generate the {programming language} code to solve it.

The code must be ready to receive input from the cli and then print the output corresponding to that input. Nothing else should be printed.

The answer should only contain the code and follow the structure:
[code]...[/code]
\end{lstlisting}
\end{minipage}\hfill
\begin{minipage}{0.45\textwidth}
\begin{lstlisting}[style=smallprompt, caption={User prompt template}, label={lst:user_prompt}]
{Extracted problem information}

{programming language} code:
\end{lstlisting}
\end{minipage}

\subsubsection{Performance Improvement Strategies}
\label{sec:prompts-perf-imp}

In addition to the standard prompt, we defined variants to support two performance-improvement strategies: \gls{ICL} and feedback incorporation.

\gls{ICL} (in-context learning) consists of augmenting the prompt with one or more example input–output pairs that illustrate the expected behavior of the solution. These examples are intended to guide the model toward producing correct outputs without modifying its internal parameters. In our setup, the system prompt template remained unchanged, while the user prompt was expanded. The expanded template, shown in Listing~\ref{lst:icl_prompt}, includes all components of the standard prompt along with one or more input–output examples.

Feedback incorporation consists of providing the model with information about why a previously generated solution failed, allowing it to refine its output in subsequent attempts. To support this prompting technique, we defined prompt variants that convey feedback based on different failure modes. We considered four outcomes: (1) the code compiled and ran but produced incorrect outputs for one or more unit tests, (2) the code failed to compile, (3) the code compiled but encountered a runtime error, and (4) the code execution exceeded the allowed time limit. Listing~\ref{lst:ut_prompt} shows how this feedback was communicated to the model in each case.

\begin{minipage}{0.35\textwidth}
\begin{lstlisting}[style=smallprompt, caption={ICL expansion of the user prompt template}, label={lst:icl_prompt}]
{Problem description}

Example:
Input: {input value}
Output: {output value}

{programming language} code:
\end{lstlisting}
\end{minipage}
\hspace{1em}
\begin{minipage}{0.55\textwidth}
\begin{lstlisting}[style=smallprompt, caption={Feedback-incorporation prompt templates}, label={lst:ut_prompt}]
% (1) The code you provided is incorrect. It compiled but did not pass the unit tests. Remember that the code should only print the output corresponding to the input.

% (2) The code you provided is incorrect. Compilation error encountered: {error message}   

% (3) The code you provided is incorrect. Runtime error encountered: {error message}   

% (4) The code took too long to run and timed out during unit tests.
\end{lstlisting}
\end{minipage}

\subsection{Procedure}
\label{sec:procedure}

In this section, we outline the key considerations across the different phases of the procedure, which are designed to ensure a rigorous and reproducible evaluation.

\subsubsection{Setup}\label{sec:proc-prep}

The first step involves defining the experimental setup:  

\begin{itemize}
    \item \textbf{Programming Language Selection}: Choose the programming languages in which the models will be evaluated.  
    \item \textbf{Unit Test and Reference Solution Thresholds}: Set the minimum and maximum thresholds for evaluation. The framework requires at least three reference solutions per language and three unit tests per problem, but these values can be adjusted to enable more thorough evaluation. Similarly, maximum thresholds can be defined to limit the number of reference solutions or unit tests used, thereby speeding up the evaluation process.
    \item \textbf{Problem Selection}: Decide whether to evaluate all problems or only a subset. If a subset is selected, establish clear criteria for selection, such as difficulty level, or problem diversity.
    \item \textbf{Problem Clustering}: Because CodeNet contains problems of varying difficulty without explicit indicators, cluster problems into difficulty levels using reproducible criteria such as prompt size or implementation complexity.  
    \item \textbf{Number of Samples per Problem}: Set the number of samples to generate per problem. A minimum of three ensures reproducibility, but higher values can improve metric estimates at the cost of longer runtimes.  
    \item \textbf{Number of Feedback Iterations}: When feedback incorporation is enabled, specify the maximum number of revision attempts after error feedback. More iterations may improve success rates but also increase runtime.  
    \item \textbf{Model Selection}: Define the models to be evaluated along with their hyperparameters.  
\end{itemize}

\subsubsection{Code Generation} 
\label{sec:proc-gen}

The Code Generation step involves presenting the instantiated prompt templates to the models and processing their responses to extract and save the generated code. In the current implementation of \texttt{PROBE}, this process is carried out using three core components, organized with Docker containers to ensure isolation and portability. The first component handles code generation. For example, this may correspond either to a third-party API for large or proprietary models or to a Docker container running Ollama for smaller models that can be executed locally. The second component, the Evaluation API, handles code evaluation by receiving the generated code along with a test and returning the outcome. The third component is a coordination script that orchestrates the workflow, managing interactions between the other components and automating the code generation process.

\subsubsection{Code Evaluation}
\label{sec:proc-eval}

The Code Evaluation step measures the correctness, proximity to reference solutions, and quality of the generated code. In the current implementation, this is achieved using a Docker container that executes the generated code and calculates all metrics for a given sample, thereby ensuring the isolation of the execution environment. The CodeBLEU score is computed using the Python \texttt{CodeBleu} library~\footnote{\url{https://pypi.org/project/codebleu/}}, while static analysis metrics are obtained with the Python implementation of the \texttt{lizard} tool~\footnote{\url{https://pypi.org/project/lizard/}}. A coordination script orchestrates the process and records the resulting metrics. All the code supporting both the code generation and evaluation is openly available at the PROBE project page\footnote{\url{https://huggingface.co/datasets/OSS-forge/PROBE}}.

\subsection{Validation \& Evolution}
\label{sec:validation}

To ensure that the benchmarking framework remains useful, trustworthy, and adaptable over time, it must be subjected to systematic validation and designed with evolution in mind. This section discusses two key aspects of this process: the benchmark properties and extensibility guidelines.

\subsubsection{Properties Analysis}
\label{sec:properties}

By adhering to core benchmarking properties, the \texttt{PROBE} benchmark ensures both reliability and practical value. These properties should be seen in line with the definitions established in problem framing (see Section~\ref{sec:problemframing}).

The \texttt{PROBE} benchmark ensures \textit{ease of use} and \textit{portability} by being fully containerized with Docker, allowing quick setup and consistent execution across different environments without requiring extensive configuration. These properties are further supported by our assumption of a black-box interaction model: the evaluated systems are accessed solely through their standard prompting interfaces, relying only on functionalities that are common across deployment environments. The framework emphasizes \textit{non-intrusiveness}: models are evaluated as-is, with prompts sent directly to the models and responses analyzed without modifying internal mechanisms. 
To guarantee \textit{repeatability}, each model is required to generate a minimum number of samples per problem (as specified in Section~\ref{sec:proc-prep}). Metrics are then computed over these samples to ensure robustness and to reduce the influence of outliers, even when full regeneration is not feasible due to the stochastic nature of \glspl{LLM}. Finally, the framework ensures \textit{representativeness} by drawing problems from real-world programming contests spanning a wide range of difficulty levels. Each problem is supported by extensive unit tests and multiple reference solutions, which enable rigorous evaluation and capture the variability of real-world coding practices. While these tasks do not encompass full-system development, they reflect the breadth and complexity of challenges relevant to practical code generation.

\subsubsection{Extensibility Guidelines}
\label{sec:extensibility}

The \texttt{PROBE} framework is designed with extensibility in mind, ensuring it remains relevant as models, tasks, and evaluation objectives evolve. Its modular structure allows extensions along several dimensions. New \textbf{metrics} can be introduced by defining the appropriate scoring functions. New \textbf{task descriptions} can also be incorporated: if unit tests are not available, they can be generated automatically using the same procedures described in Section~\ref{sec:workload-unit}. However, to evaluate proximity to valid solutions and code quality in addition to functional correctness, reference solutions must be provided if they are not already available. The same requirement applies when introducing additional \textbf{programming languages}: while functional correctness can be assessed with unit tests alone, evaluation of proximity to valid solutions and code quality depends on the presence of suitable reference solutions, either in the dataset or supplied by the user. Finally, alternative \textbf{prompting strategies}, such as chain-of-thought prompting, can be incorporated by modifying or adding new prompt templates. This modularity ensures that the benchmark remains flexible, rigorous, and adaptable to future advances in \gls{LLM} capabilities and evaluation needs.

\section{Experimental Setup}
\label{sec:experimental}

This section presents our experimental study, detailing the setup parameters.

\paragraph{Programming Language Selection:}  
Our evaluation spans five programming languages: Python, C++, C, Java, and Rust. These languages were selected to capture a broad spectrum of syntactic complexity, semantic rigor, and real-world usage. Python offers a high-level, natural-language-like syntax that minimizes compilation issues. C++ and Java occupy a middle ground, with stricter typing and more elaborate structure, while C and Rust are among the most demanding due to low-level control and rigorous safety constraints.

Rust, in particular, was included because it remains comparatively underrepresented in publicly available training data, providing a meaningful test case for evaluating model performance under limited exposure. However, this limited ecosystem also means that some automated evaluation tools, most notably CodeBLEU, lack full support for Rust. Despite this limitation, we opted to retain Rust in the benchmark due to its conceptual and practical importance, while excluding it specifically from analyses that rely on CodeBLEU’s proximity-to-solution metrics. This choice allows us to preserve the value of Rust as a challenging and informative test language without compromising the consistency of those tool-dependent evaluations.

\paragraph{Unit Test and Reference Solution Thresholds:}  
The minimum thresholds are kept at the framework’s required values of three unit tests per problem and three reference solutions per language. In addition, we set a maximum of ten unit tests per problem and 250 reference solutions per problem and language, chosen randomly. This ensures a balanced workload across problems while keeping execution time manageable.  

\paragraph{Problem Selection:}  
To ensure robust evaluation, we retained only problems with at least three reference solutions in both Python and C++, the most common languages in CodeNet. This resulted in 1,651 problems, which balances breadth of coverage with reliability of metrics. This allows proximity and code quality metrics to be computed in all problems in at least two languages. Following the selection process, the final dataset, whose characteristics are summarized in Table~\ref{tab:dataset-final}, comprised 1,651 problems. Each problem had between three and ten unit tests, with an average of 8.55, which is still above several widely used benchmarks~\citep{chen2021evaluatinglargelanguagemodels, austin2021programsynthesislargelanguage}. In addition, every problem had between three and 250 reference solutions in Python and C++, fully satisfying the requirements for evaluating the proximity to a valid solution and the code quality. For Java, C, and Rust, reference solutions ranged from 0 to 250, but some problems fell short of the minimum threshold: 83 in Java, 302 in C, and 349 in Rust; for these problems, proximity and quality metrics will not be computed.

\paragraph{Problem Clustering:} 
To enable difficulty-based analysis, problems were grouped into levels using clustering over static code metrics derived from Python reference solutions. Specifically, for each problem, we computed the average cyclomatic complexity, effort, and \gls{SLOC} across all validated solutions. Prior to clustering, all features were standardized using z-score normalization to ensure comparable scale and equal contribution of each metric.

We evaluated multiple clustering approaches, including $k$-means, DBSCAN, HDBSCAN, and Gaussian Mixture Models, with empirically tuned hyperparameters. Cluster quality was assessed primarily through visual inspection of cluster separation and distribution balance. Among the evaluated methods, $k$-means with $k=5$ produced the clearest separation while also yielding the most balanced cluster sizes. The clustering process initially produced five groups; however, the two highest-difficulty clusters were merged because one cluster contained only seven problems, which is insufficient for meaningful statistical analysis, resulting in four final difficulty levels.

We recognize that these metrics capture implementation complexity rather than intrinsic problem difficulty. To mitigate this limitation, metrics are averaged across all existing reference solutions, reducing the influence of atypical implementations. Furthermore, the resulting clustering was validated empirically: model performance decreases consistently as difficulty increases, across multiple models and programming languages, indicating that the derived levels capture meaningful differences for analysis purposes.

Only Python solutions were used due to tool support and consistency. While this introduces a limitation, the observed cross-language performance trends suggest that the resulting difficulty estimates generalize beyond a single programming language. The full procedure is reproducible, ensuring consistent results across experimental runs.

\paragraph{Number of Samples per Problem:}  
We opted for the generation of five samples per problem to balance obtaining reliable estimates of evaluation metrics and keeping the runtime manageable.

\paragraph{Number of Feedback Iterations:}  
We selected two feedback iterations. This choice reflects a trade-off between allowing the models enough opportunities to correct their mistakes after receiving execution feedback and maintaining a practical evaluation runtime.

\paragraph{Model Selection:}  
Table~\ref{tab:model_selection} lists the models evaluated. We evaluated two large, commercial models, GPT-4.1-mini and Gemini-2.0-flash, and four smaller open-source models: Qwen2.5:14b, Qwen2.5-coder:7b, Qwen2.5-coder:14b, and Deepseek-coder-v2. Among the latter, we cover both general-purpose and code-specialized models, as well as standard dense and \gls{MoE} architectures. Multiple models of the Qwen2.5-coder family with different parameter counts are also included. This diverse selection enables us to investigate how architecture, training data, and parameter size impact performance, while providing confidence that the results generalize beyond the models explicitly tested. Hyperparameters were fixed for all models to ensure consistency and comparability across evaluations. Temperature was set to 0.6 (as prior work~\citep{chen2021evaluatinglargelanguagemodels} shows this value performs well when generating multiple samples per problem), while top-p was set to 1 and top-k to 50. Although per-model hyperparameter tuning could improve individual performance, it was not feasible at our evaluation scale and would introduce additional variability, potentially affecting fairness. Instead, we adopt a shared configuration to provide a controlled and reproducible comparison across models. Furthermore, our primary objective is to evaluate the proposed benchmarking methodology rather than to maximize the performance of individual models.

\begin{table}[t]
    \centering
    \caption{Experimental Settings}
    
    \begin{subtable}[t]{1.0\linewidth}
        \centering
        \caption{Final Dataset Characteristics} 
        \label{tab:dataset-final}
        \begin{tabular}{r|l}
            \textbf{Languages} & Python, C++, Java, C, Rust \\
            \textbf{Problems} & 1651 \\
            \textbf{Prob. diff.} & \textbf{0:} 796, \textbf{1:} 520, \textbf{2:} 261, \textbf{3:} 74 \\
            \textbf{Unit Tests} & 3-10 \\
            \textbf{Ref. Sols} & \makecell[l]{3-250 (Python, C++) \\ 0-250 (Java, C) \\ 0-180 (Rust)}
        \end{tabular}
    \end{subtable}
    \vspace{1em}
    \begin{subtable}[t]{1.0\linewidth}
        \centering
        \caption{Models Selected} 
        \label{tab:model_selection}
        \begin{tabular}{c|cc}
            \textbf{Model} & \textbf{Arch.} & \textbf{Purpose} \\ \hline
            GPT-4.1-mini            & Dense      & General \\
            Gemini-2.0-flash         & Dense     & General \\
            Qwen2.5:14b              & Dense     & General \\
            Qwen2.5-Coder:7b         & Dense     & Code \\
            Qwen2.5-Coder:14b        & Dense     & Code \\
            Deepseek-coder-v2         & \gls{MoE} & Code
        \end{tabular}
    \end{subtable}

\end{table}

\section{Results}
\label{sec:results}

In this section, we first examine model performance along each of the three evaluation dimensions introduced in \texttt{PROBE}: functional correctness (Section~\ref{sec:results-functional}), proximity to a valid solution (Section~\ref{sec:results-proximity}), and code quality (Section~\ref{sec:results-quality}). For each dimension, we begin with the baseline prompting setup and then assess the effects of \gls{ICL}, feedback incorporation, and task difficulty. Finally, we provide an analysis of the most common errors observed in the generated code, accompanied by illustrative examples (Section~\ref{sec:results-common-problems}).

\subsection{Functional Correctness}
\label{sec:results-functional}

\subsubsection{Baseline}

\begin{figure*}
    \centering
    \includegraphics[width=1.0\linewidth]{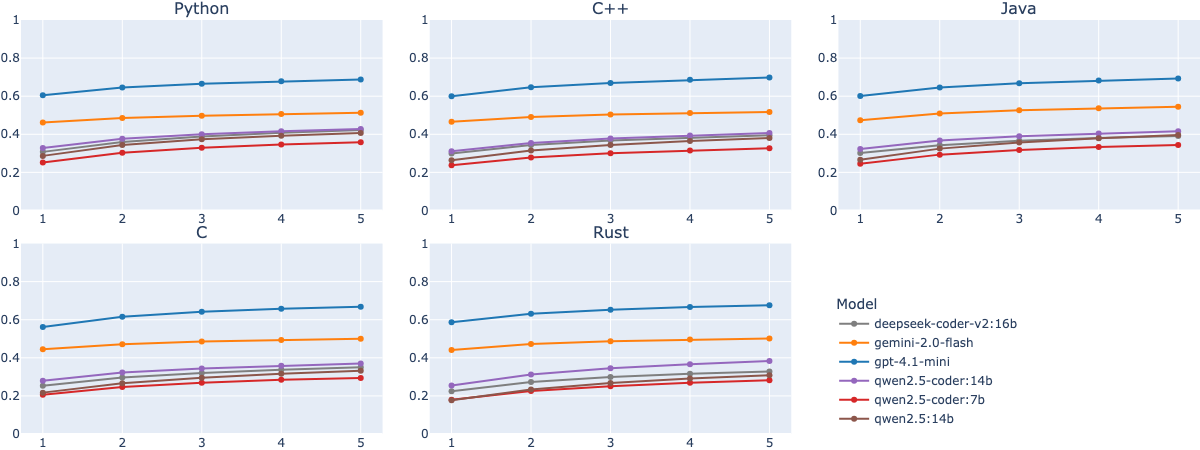}
    \caption{Pass@k values for the different models and languages}
    \label{fig:passk}
\end{figure*}

Figure~\ref{fig:passk} presents the pass@k results for all models and programming languages. The figure illustrates a clear relationship between model size and performance. The larger proprietary models achieve higher pass@k values across every language, and this advantage is stable throughout the entire range of k. The same pattern appears, at a smaller scale, within the open-source models: qwen2.5-coder:7b consistently attains the lowest pass@k values in every language, while the 14b variants perform noticeably better. Despite these differences, the strongest models remain far from solving all tasks, as none exceed a pass@k of 0.7 in any language. This indicates that while scale improves code generation quality, it does not eliminate a substantial set of problems that remain unsolved even with multiple attempts.

Comparing pass@1 with pass@5 shows that additional samples provide a modest but steady improvement across all models. The gain from increasing k is relatively small, typically no more than 0.1-0.12, which suggests that the generated code is not extremely diverse, since drawing more samples does not significantly alter the model's performance.

The results also reveal that programming language characteristics have a clear influence on model performance. Python consistently yields the highest pass rates, followed by C++ and Java at similar levels, then C, with Rust emerging as the most challenging language. Although this ranking remains the same for every model, from the smallest open-source systems to the strongest proprietary ones, the gap between Rust and the other languages grows substantially in the smaller models. A plausible explanation is that Rust has comparatively fewer resources and examples in the training data, which makes it harder for lower-capacity models to acquire the necessary abstractions. The larger models appear better able to generalize to Rust’s stricter semantics and more complex constraints, whereas the smaller ones struggle to do so reliably.

Table~\ref{tab:outcome} summarizes the distribution of outcomes across all models and programming languages. The proportion of \textit{No Code} events is extremely small for every model and language. This indicates that all systems consistently attempt to produce code when prompted. In practice, this means that most failures do not stem from refusal or inability to output code, but rather from the quality or correctness of the generated programs.

Once we consider only the cases where code was produced, the four possible failure types reveal clear language-dependent patterns. In Python, most incorrect outputs fall into the \textit{Failed} or \textit{Timeout} categories, as expected given Python’s interpreted nature and its more flexible syntax. In contrast, compiled languages exhibit a very different profile. A substantial portion of errors originates from code that does not compile, reflecting the stricter syntactic and type constraints of languages such as C, C++, and especially Rust. Rust shows the highest proportion of non-compilation outcomes. A plausible explanation is the relatively limited representation of Rust in current model training data, combined with its strict syntax, borrowing rules, and ownership model. As a result, models more frequently generate code with small inaccuracies that prevent compilation. This also has a direct impact on the other outcomes, as Rust is the language with the fewest unit tests that fail at execution time, simply because many generated programs never reach the point of being run against the tests.

There are also clear differences across models. The larger proprietary systems, GPT-4.1-mini and Gemini-2.0-flash, consistently produce the lowest rates of compilation failures and runtime errors. GPT-4.1-mini also maintains the lowest timeout rate across all languages. This shows that their scale and training breadth enable a more robust handling of language syntax and program structure. Among the open-source models, the general-purpose Qwen-2.5:14b shows noticeably higher rates of \texttt{No compile} compared to its code-specialized variants, suggesting that code-specific fine-tuning plays an important role in helping models internalize the syntactic and structural requirements of programming languages.

Finally, when looking at the rates for \enquote{passed}, we can see that the results match what was observed for pass@k. Although the rates are very similar, the highest values can be observed in Python, with a gradual decline in Java, C++, and C, and with the lowest values observed in Rust. This trend aligns with the relative complexity of each language. Dynamic languages such as Python are more forgiving of minor inaccuracies, while compiled languages demand strict syntactic and structural correctness.

\begin{table}[t]
    \centering
    \caption{Overall Outcome Rate values}
    \label{tab:outcome}
    \begin{tabular}{llrrrrrr}
    \toprule
    Model & Lang. & Pass. & Fail. & Time. & Runtime. & Comp. & No code \\
    \midrule
    \rowcolor{gray!15}
    deepseek-coder-v2 & Python & 0.41 & 0.34 & 0.14 & 0.11 & 0.00 & 0.00 \\
    \rowcolor{gray!15}
    gemini-2.0-flash & Python & 0.57 & 0.18 & 0.20 & 0.04 & 0.00 & 0.00 \\
    \rowcolor{gray!15}
    gpt-4.1-mini & Python & 0.70 & 0.23 & 0.05 & 0.02 & 0.00 & 0.00 \\
    \rowcolor{gray!15}
    qwen2.5-coder:14b & Python & 0.44 & 0.37 & 0.11 & 0.08 & 0.00 & 0.00 \\
    \rowcolor{gray!15}
    qwen2.5-coder:7b & Python & 0.36 & 0.38 & 0.12 & 0.13 & 0.01 & 0.00 \\
    \rowcolor{gray!15}
    qwen2.5:14b & Python & 0.38 & 0.34 & 0.10 & 0.17 & 0.01 & 0.00 \\ \hline
    deepseek-coder-v2 & C++ & 0.40 & 0.37 & 0.10 & 0.02 & 0.11 & 0.00 \\
    gemini-2.0-flash & C++ & 0.57 & 0.20 & 0.15 & 0.01 & 0.07 & 0.00 \\
    gpt-4.1-mini & C++ & 0.68 & 0.25 & 0.03 & 0.01 & 0.03 & 0.00 \\
    qwen2.5-coder:14b & C++ & 0.42 & 0.38 & 0.07 & 0.02 & 0.11 & 0.00 \\
    qwen2.5-coder:7b & C++ & 0.34 & 0.42 & 0.08 & 0.03 & 0.13 & 0.00 \\
    qwen2.5:14b & C++ & 0.35 & 0.32 & 0.05 & 0.03 & 0.26 & 0.00 \\ \hline
    \rowcolor{gray!15}
    deepseek-coder-v2 & Java & 0.41 & 0.38 & 0.10 & 0.07 & 0.04 & 0.00 \\
    \rowcolor{gray!15}
    gemini-2.0-flash & Java & 0.59 & 0.23 & 0.13 & 0.03 & 0.02 & 0.00 \\
    \rowcolor{gray!15}
    gpt-4.1-mini & Java & 0.69 & 0.24 & 0.03 & 0.03 & 0.02 & 0.00 \\
    \rowcolor{gray!15}
    qwen2.5-coder:14b & Java & 0.43 & 0.37 & 0.08 & 0.05 & 0.07 & 0.00 \\
    \rowcolor{gray!15}
    qwen2.5-coder:7b & Java & 0.36 & 0.40 & 0.08 & 0.08 & 0.07 & 0.00 \\
    \rowcolor{gray!15}
    qwen2.5:14b & Java & 0.35 & 0.33 & 0.05 & 0.09 & 0.18 & 0.00 \\ \hline
    deepseek-coder-v2 & C & 0.37 & 0.40 & 0.08 & 0.05 & 0.09 & 0.00 \\
    gemini-2.0-flash & C & 0.56 & 0.24 & 0.11 & 0.05 & 0.03 & 0.00 \\
    gpt-4.1-mini & C & 0.65 & 0.25 & 0.02 & 0.01 & 0.07 & 0.00 \\
    qwen2.5-coder:14b & C & 0.39 & 0.41 & 0.06 & 0.04 & 0.10 & 0.00 \\
    qwen2.5-coder:7b & C & 0.31 & 0.41 & 0.07 & 0.05 & 0.16 & 0.00 \\
    qwen2.5:14b & C & 0.30 & 0.36 & 0.04 & 0.05 & 0.25 & 0.00 \\ \hline
    \rowcolor{gray!15}
    deepseek-coder-v2 & Rust & 0.31 & 0.26 & 0.08 & 0.11 & 0.24 & 0.00 \\
    \rowcolor{gray!15}
    gemini-2.0-flash & Rust & 0.55 & 0.18 & 0.14 & 0.07 & 0.06 & 0.00 \\
    \rowcolor{gray!15}
    gpt-4.1-mini & Rust & 0.67 & 0.21 & 0.03 & 0.03 & 0.06 & 0.00 \\
    \rowcolor{gray!15}
    qwen2.5-coder:14b & Rust & 0.34 & 0.24 & 0.07 & 0.07 & 0.29 & 0.00 \\
    \rowcolor{gray!15}
    qwen2.5-coder:7b & Rust & 0.25 & 0.24 & 0.07 & 0.13 & 0.31 & 0.00 \\
    \rowcolor{gray!15}
    qwen2.5:14b & Rust & 0.24 & 0.17 & 0.04 & 0.13 & 0.42 & 0.00 \\
    \bottomrule
    \end{tabular}
\end{table}

\subsubsection{In-Context Learning}

Table \ref{tab:icl-func} shows the impact of 1-shot learning on the functional correctness metrics per language and per model. As we can see, this technique had next to no impact on performance across the board. For languages (Table \ref{tab:icl-func-lang}), the changes in pass@1 and pass@5 are effectively zero, with all values ranging between –0.00 and 0.01. The largest improvements (still marginal) appear in Java and Python, each showing a +0.01 increase in pass@1, while other languages remain unchanged. Similarly, the distribution of outcome categories (Pass., Fail., Time., Error, Comp., No Code) varies only within $\pm$ 0.02, with the largest deviation being a tiny –0.02 reduction in failed (incorrect output) C submissions.

The per-model results (Table \ref{tab:icl-func-model}) reveal the same trend. All models show changes of at most $\pm$ 0.02, with gpt-4.1-mini exhibiting the largest (yet still negligible) gain in pass@5 (+0.02) and pass rate (+0.02). Other models fluctuate by only $\pm$ 0.01 in any category, and several (e.g., deepseek-coder-v2, qwen2.5-coder:7b) show virtually no differences at all.

These results can be explained by the fact that, unlike what happens in other benchmarks~\citep{chen2021evaluatinglargelanguagemodels}, our base prompt already contains detailed instructions about the input and output formats. This means that including a concrete example provides little new information for the models, leading to minimal measurable benefit from 1-shot \gls{ICL}.

\begin{table}[t]
\centering
\caption{Impact of \gls{ICL} on the functional correctness metrics}
\label{tab:icl-func}

    \begin{subtable}[t]{1.0\textwidth}
        \setlength{\tabcolsep}{4pt}
        \centering
        \caption{Per-language impact}
        \label{tab:icl-func-lang}
        \begin{tabular}{l|rr|rrrrrr}
        \toprule
        Language & pass@1 & pass@5 & Pass. & Fail. & Time. & Error & Comp. & No Code \\
        \midrule
        C & -0.00 & -0.00 & -0.00 & -0.02 & -0.00 & -0.00 & 0.02 & 0.00 \\
        C++ & 0.00 & 0.00 & 0.01 & -0.00 & -0.00 & -0.00 & -0.00 & 0.00 \\
        Java & 0.01 & 0.01 & 0.01 & -0.00 & -0.00 & -0.01 & -0.00 & 0.00 \\
        Python & 0.01 & 0.00 & 0.01 & -0.00 & -0.00 & -0.00 & 0.00 & 0.00 \\
        Rust & 0.00 & 0.00 & 0.01 & -0.00 & 0.00 & -0.00 & -0.00 & 0.00 \\
        \hline
        Average & 0.00 & 0.00 & 0.01 & -0.01 & -0.00 & -0.00 & 0.00 & 0.00 \\
        \bottomrule
        \end{tabular}
    \end{subtable}

    \vspace{1em}

    \begin{subtable}[t]{1.0\textwidth}
        \setlength{\tabcolsep}{4pt}
        \centering
        \caption{Per-model impact}
        \label{tab:icl-func-model}
        \begin{tabular}{l|rr|rrrrrr}
        \toprule
        Model & pass@1 & pass@5 & Pass. & Fail. & Time. & Error & Comp. & No Code \\
        \midrule
        deepseek-coder-v2 & 0.00 & -0.00 & 0.00 & 0.00 & -0.00 & -0.00 & 0.00 & 0.00 \\
        gemini-2.0-flash & -0.00 & -0.00 & 0.00 & -0.01 & 0.00 & -0.00 & 0.01 & 0.00 \\
        gpt-4.1-mini & 0.01 & 0.02 & 0.02 & -0.01 & 0.00 & -0.00 & -0.00 & 0.00 \\
        qwen2.5-coder:14b & 0.00 & 0.00 & 0.01 & -0.01 & -0.00 & -0.00 & 0.01 & 0.00 \\
        qwen2.5-coder:7b & 0.00 & 0.00 & 0.01 & -0.00 & -0.00 & -0.00 & -0.00 & 0.00 \\
        qwen2.5:14b & 0.00 & -0.00 & 0.00 & -0.01 & -0.00 & -0.01 & 0.01 & 0.00 \\
        \hline
        Average & 0.00 & 0.00 & 0.01 & -0.01 & -0.00 & -0.00 & 0.00 & 0.00 \\
        \bottomrule
        \end{tabular}
    \end{subtable}

\end{table}

\subsubsection{Feedback-incorporation}

\begin{table}[t]
    \centering
    \caption{Impact of feedback on the functional correctness metrics}
    \label{tab:feedback-func}

    \begin{subtable}[t]{1.0\textwidth}
        \setlength{\tabcolsep}{4pt}
        \centering
        \caption{Per-language impact}
        \label{tab:feedback-func-language}    
        \begin{tabular}{l|rr|rrrrrr}
        \toprule
        Language & pass@1 & pass@5 & Pass. & Fail. & Time. & Error & Comp. & No Code \\
        \midrule
        C & 0.05 & 0.05 & 0.06 & -0.00 & -0.01 & 0.00 & -0.04 & 0.00 \\
        C++ & 0.05 & 0.06 & 0.06 & 0.00 & -0.02 & 0.00 & -0.05 & 0.00 \\
        Java & 0.04 & 0.04 & 0.04 & -0.00 & -0.02 & -0.01 & -0.01 & 0.00 \\
        Python & 0.04 & 0.05 & 0.04 & 0.01 & -0.03 & -0.02 & -0.00 & 0.00 \\
        Rust & 0.06 & 0.06 & 0.07 & 0.03 & -0.01 & -0.01 & -0.08 & 0.00 \\
        \hline
        Average & 0.05 & 0.05 & 0.05 & 0.01 & -0.02 & -0.01 & -0.04 & 0.00 \\
        \bottomrule
        \end{tabular}
    \end{subtable}

    \vspace{1em}

    \begin{subtable}[t]{1.0\textwidth}
        \setlength{\tabcolsep}{4pt}
        \centering
        \caption{Per-model impact}
        \label{tab:feedback-func-model}
        \begin{tabular}{l|rr|rrrrrr}
        \toprule
        Model & pass@1 & pass@5 & Pass. & Fail. & Time. & Error & Comp. & No Code \\
        \midrule
        deepseek-coder-v2 & 0.05 & 0.06 & 0.05 & -0.01 & -0.02 & -0.01 & -0.02 & 0.00 \\
        gemini-2.0-flash & 0.05 & 0.05 & 0.05 & 0.00 & -0.03 & -0.01 & -0.02 & 0.00 \\
        gpt-4.1-mini & 0.04 & 0.03 & 0.04 & -0.01 & -0.01 & -0.01 & -0.02 & 0.00 \\
        qwen2.5-coder:14b & 0.04 & 0.05 & 0.05 & 0.02 & -0.02 & -0.00 & -0.05 & 0.00 \\
        qwen2.5-coder:7b & 0.04 & 0.04 & 0.04 & 0.02 & -0.03 & -0.01 & -0.03 & 0.00 \\
        qwen2.5:14b & 0.07 & 0.07 & 0.08 & 0.02 & -0.01 & -0.01 & -0.08 & 0.00 \\
        \hline
        Average & 0.05 & 0.05 & 0.05 & 0.01 & -0.02 & -0.01 & -0.04 & 0.00 \\
        \bottomrule
        \end{tabular}
    \end{subtable}
        
\end{table}

Unlike \gls{ICL}, feedback incorporation had a clear and consistent impact on model performance (Table~\ref{tab:feedback-func}). Across languages (Table~\ref{tab:feedback-func-language}), we observe noticeable gains in both pass@1 and pass@5, with average improvements of +0.05 for both metrics. The increase in the outcome rate for outcome \enquote{passed} (+0.05 on average) aligns with this trend. These gains come primarily from a substantial reduction in compilation errors (–0.04 on average), while other failure modes such as runtime errors, timeouts, or incorrect outputs have smaller shifts. This is expected: compilation errors are often straightforward to fix once explicit feedback is provided (e.g., missing imports, undeclared variables, simple type mismatches), whereas diagnosing runtime issues or incorrect algorithmic logic is considerably more challenging and the errors messages may be less informative (e.g. segmentation faults).

The reduction in compilation errors is particularly pronounced in C (–0.04), C++ (–0.05), and Rust (–0.08). This pattern makes sense given that these languages have stricter compilation pipelines and more verbose error messages, meaning that small syntactic or type-related issues in the initial generation lead to disproportionately high rates of compilation failures. Feedback therefore provides highly actionable information for correcting these initial errors. Rust shows the largest improvement, which can likely be attributed to its weaker initial performance compared to the other languages and the especially high number of compilation errors produced in the first generation. Because Rust’s compile-time diagnostics are detailed and explicit, the models can leverage this feedback effectively even if their initial attempts are less accurate.

A similar trend appears across models (Table~\ref{tab:feedback-func-model}). All systems benefit from feedback, with average increases of +0.05 in pass@1 and pass@5 and a corresponding +0.05 improvement in the proportion of correct unit tests. Again, these gains are driven by a reduction in compilation failures (–0.04), while changes in runtime failures, timeouts, and logical errors remain small.

Among the evaluated models, qwen2.5:14b exhibits the most significant improvement, with +0.07 in pass@1, +0.07 in pass@5, and +0.08 in the outcome rate for \enquote{passed}. This can be partly explained by its comparatively weaker initial performance, which means there is more room for improvement after receiving explicit feedback. Additionally, because qwen2.5:14b is a general-purpose model rather than a coding-specialized one, it may benefit more from the clarity of compiler messages, leveraging this external structure to compensate for its less specialized code-generation capabilities. Its coding-oriented variants already make fewer low-level mistakes initially, leaving less opportunity for feedback to produce dramatic changes.

Overall, these results show that incorporating compiler feedback is substantially more effective than providing an in-context example, especially in languages and models with higher initial rates of simple, fixable compilation errors.

\subsubsection{Problem Difficulty}

\begin{table}[t]
\centering
\caption{pass@k performance across difficulty levels}
\label{tab:passk-diff}

    \begin{subtable}[t]{0.25\textwidth}
        \centering
        \vspace{1.8em}
        \begin{tabular}{l}
        \toprule
        Diff. Level \\
        \midrule
        deepseek-coder-v2 \\
        gemini-2.0-flash \\
        gpt-4.1-mini \\
        qwen2.5-coder:14b \\
        qwen2.5-coder:7b \\
        qwen2.5:14b \\
        \bottomrule
        \end{tabular}
    \end{subtable}
    \hfill
    \begin{subtable}[t]{0.35\textwidth}
        \centering
        \caption{pass@1}
        \begin{tabular}{rrrr}
        \toprule
        0 & 1 & 2 & 3 \\
        \midrule
        0.46 & 0.14 & 0.04 & 0.06 \\
        0.69 & 0.31 & 0.14 & 0.12 \\
        0.77 & 0.51 & 0.31 & 0.20 \\
        0.51 & 0.14 & 0.05 & 0.05 \\
        0.39 & 0.09 & 0.02 & 0.03 \\
        0.44 & 0.09 & 0.02 & 0.02 \\
        \bottomrule
        \end{tabular}
    \end{subtable}
    \hfill
    \begin{subtable}[t]{0.35\textwidth}
        \centering
        \caption{pass@5}
        \begin{tabular}{rrrr}
        \toprule
        0 & 1 & 2 & 3 \\
        \midrule
        0.60 & 0.23 & 0.09 & 0.09 \\
        0.73 & 0.39 & 0.21 & 0.16 \\
        0.84 & 0.63 & 0.43 & 0.31 \\
        0.64 & 0.23 & 0.10 & 0.11 \\
        0.53 & 0.17 & 0.05 & 0.08 \\
        0.61 & 0.18 & 0.07 & 0.06 \\
        \bottomrule
        \end{tabular}
    \end{subtable}

\end{table}

Across the different problem difficulty levels, we observe consistent performance degradation as tasks become more challenging, with clear patterns in both pass@k and outcome distributions (Tables \ref{tab:passk-diff} and \ref{tab:outcomeRate-diff}). It is important to note that, although we only report the per-model results to avoid excessive table volume, the same trends can be seen in the per-language breakdown.

Starting with pass@k (Table~\ref{tab:passk-diff}), the drop in pass@1 and pass@5 between difficulty level 0 and level 3 is substantial for every model, reflecting the increasing algorithmic and structural complexity of the benchmark. At difficulty level 0, most systems achieve moderate-to-high pass@1 scores (0.39–0.77), but these fall sharply by level 3, where values remain around 0.02–0.20. The same pattern appears in pass@5: while scores range from 0.53 to 0.84 at level 0, they drop to only 0.06–0.31 for level 3 tasks.

\begin{table}[t]
    \centering
    \caption{OutcomeRate performance across difficulty levels}
    \label{tab:outcomeRate-diff}

    \begin{subtable}[t]{0.25\textwidth}
        \centering
        \vspace{1.8em}
        \begin{tabular}{l}
        \toprule
        Diff. Level \\
        \midrule
        deepseek-coder-v2 \\
        gemini-2.0-flash \\
        gpt-4.1-mini \\
        qwen2.5-coder:14b \\
        qwen2.5-coder:7b \\
        qwen2.5:14b \\
        \bottomrule
        \end{tabular}
    \end{subtable}
    \hfill
    \begin{subtable}[t]{0.35\textwidth}
        \centering
        \caption{Passed}
        \begin{tabular}{rrrr}
        \toprule
        0 & 1 & 2 & 3 \\
        \midrule
        0.56 & 0.24 & 0.15 & 0.15 \\
        0.77 & 0.44 & 0.30 & 0.27 \\
        0.83 & 0.61 & 0.45 & 0.37 \\
        0.61 & 0.25 & 0.16 & 0.17 \\
        0.50 & 0.18 & 0.12 & 0.12 \\
        0.53 & 0.16 & 0.09 & 0.07 \\
        \bottomrule
        \end{tabular}
    \end{subtable}
    \hfill
    \begin{subtable}[t]{0.35\textwidth}
        \centering
        \caption{Failed}
        \begin{tabular}{rrrr}
        \toprule
        0 & 1 & 2 & 3 \\
        \midrule
        0.29 & 0.42 & 0.40 & 0.39 \\
        0.13 & 0.26 & 0.29 & 0.32 \\
        0.14 & 0.29 & 0.38 & 0.42 \\
        0.27 & 0.45 & 0.40 & 0.43 \\
        0.32 & 0.44 & 0.38 & 0.38 \\
        0.26 & 0.37 & 0.31 & 0.29 \\
        \bottomrule
        \end{tabular}
    \end{subtable}

    \vspace{1em}
    
    \begin{subtable}[t]{0.25\textwidth}
        \centering
        \vspace{1.8em}
        \begin{tabular}{l}
        \toprule
        Diff. Level \\
        \midrule
        deepseek-coder-v2 \\
        gemini-2.0-flash \\
        gpt-4.1-mini \\
        qwen2.5-coder:14b \\
        qwen2.5-coder:7b \\
        qwen2.5:14b \\
        \bottomrule
        \end{tabular}
    \end{subtable}
    \hfill
    \begin{subtable}[t]{0.35\textwidth}
        \centering
        \caption{Timeout}
        \begin{tabular}{rrrr}
        \toprule
        0 & 1 & 2 & 3 \\
        \midrule
        0.06 & 0.13 & 0.16 & 0.14 \\
        0.07 & 0.20 & 0.27 & 0.24 \\
        0.01 & 0.04 & 0.06 & 0.06 \\
        0.04 & 0.10 & 0.14 & 0.12 \\
        0.05 & 0.11 & 0.14 & 0.12 \\
        0.03 & 0.08 & 0.09 & 0.09 \\
        \bottomrule
        \end{tabular}
    \end{subtable}
    \hfill
    \begin{subtable}[t]{0.35\textwidth}
        \centering
        \caption{Runtime Error}
        \begin{tabular}{rrrr}
        \toprule
        0 & 1 & 2 & 3 \\
        \midrule
        0.04 & 0.09 & 0.13 & 0.13 \\
        0.01 & 0.05 & 0.08 & 0.07 \\
        0.01 & 0.02 & 0.03 & 0.04 \\
        0.03 & 0.06 & 0.10 & 0.08 \\
        0.05 & 0.10 & 0.14 & 0.14 \\
        0.06 & 0.12 & 0.15 & 0.16 \\
        \bottomrule
        \end{tabular}
    \end{subtable}

    \vspace{1em}
    
    \begin{subtable}[t]{0.25\textwidth}
        \centering
        \vspace{1.8em}
        \begin{tabular}{l}
        \toprule
        Diff. Level \\
        \midrule
        deepseek-coder-v2 \\
        gemini-2.0-flash \\
        gpt-4.1-mini \\
        qwen2.5-coder:14b \\
        qwen2.5-coder:7b \\
        qwen2.5:14b \\
        \bottomrule
        \end{tabular}
    \end{subtable}
    \hfill
    \begin{subtable}[t]{0.35\textwidth}
        \centering
        \caption{Compilation Error}
        \begin{tabular}{rrrr}
        \toprule
        0 & 1 & 2 & 3 \\
        \midrule
        0.05 & 0.11 & 0.17 & 0.19 \\
        0.02 & 0.05 & 0.07 & 0.10 \\
        0.02 & 0.04 & 0.08 & 0.10 \\
        0.06 & 0.14 & 0.21 & 0.21 \\
        0.08 & 0.16 & 0.22 & 0.25 \\
        0.13 & 0.27 & 0.35 & 0.39 \\
        \bottomrule
        \end{tabular}
    \end{subtable}
    \hfill
    \begin{subtable}[t]{0.35\textwidth}
        \centering
        \caption{No Code}
        \begin{tabular}{rrrr}
        \toprule
        0 & 1 & 2 & 3 \\
        \midrule
        0.04 & 0.09 & 0.13 & 0.13 \\
        0.01 & 0.05 & 0.08 & 0.07 \\
        0.01 & 0.02 & 0.03 & 0.04 \\
        0.03 & 0.06 & 0.10 & 0.08 \\
        0.05 & 0.10 & 0.14 & 0.14 \\
        0.06 & 0.12 & 0.15 & 0.16 \\
        \bottomrule
        \end{tabular}
    \end{subtable}
    
\end{table}

Larger models degrade more slowly, but even they struggle with the hardest problems. For example, gpt-4.1-mini, the strongest model overall, shows the strongest performance in the hardest problems, yet still reaches only 0.20 pass@1 on difficulty 3, meaning that when asked for a single solution, four out of five answers remain incorrect. This highlights the substantial jump in complexity introduced by the harder tasks and shows that sheer scale does not guarantee reliable single-shot correctness.

In contrast, smaller models such as qwen2.5:14b and qwen2.5-coder:7b exhibit the steepest declines. Interestingly, although qwen2.5:14b consistently outperforms qwen2.5-coder:7b on easier tasks (e.g., pass@1 of 0.44 vs. 0.39 on level 0), this relationship reverses at higher difficulty. At difficulty 3, qwen2.5-coder:7b achieves a higher pass@1 (0.03 vs. 0.02) and pass@5 (0.08 vs. 0.06). This suggests that code-specialized fine-tuning confers advantages specifically for more complex tasks, where syntactic discipline, code structure, and familiarity with common programming patterns become increasingly important.

The outcome rate breakdown (Table \ref{tab:outcomeRate-diff}) reinforces these trends. At difficulty level 0, the proportion of passed unit tests is relatively high for all models (0.50–0.83). As difficulty increases, the \enquote{passed} rate steadily decreases, ultimately dropping to 0.07–0.37 at level 3. The inverse holds for failure modes: incorrect outputs, timeouts, runtime errors, and compilation errors all increase as problem complexity grows.

Despite this consistent rise, the error distributions reveal interesting model-specific behaviors. For instance, Gemini-2.0-Flash produces a disproportionately high number of timeouts at higher difficulties, rising from 0.07 at level 0 to 0.24 at level 3, by far the highest among all evaluated models. Compilation errors also grow more sharply for models with weaker syntactic reliability. For example, qwen2.5:14b sees its compilation error rate nearly triple, from 0.13 at level 0 to 0.39 at level 3. This suggests that more complex prompts amplify issues in parsing and type handling, particularly in stricter languages, where small inaccuracies can cascade into full compilation failures.

The strongest models show relatively smaller increases in low-level errors. Gpt-4.1-mini, for example, increases from 0.02 to 0.10 compilation errors between difficulty levels 0 and 3. However, its advantage is less pronounced for other error types: for the hardest problems, gpt-4.1-mini remains one of the models with the highest rates of incorrect outputs (0.42 at difficulty 3), and Gemini-2.0-Flash consistently produces the most timeouts. Taken together, these patterns indicate that while larger or more capable models maintain structural and syntactic coherence even under challenging conditions, smaller or general-purpose systems degrade across all error categories.

Overall, these results show that performance drops sharply as problem difficulty increases, highlighting that \glspl{LLM} are still limited when it comes to solving harder tasks, and that how they fail is strongly model-specific, reflecting differences in training data, specialization, and architectural capacity.

\begin{tcolorbox}[colback=gray!10, colframe=black, title=\textbf{Key Takeaways}]
\begin{itemize}
    \item \textbf{Model scale strongly correlates with performance}, but even the best models remain far from reliable (pass@1 $\leq$ 0.70).
    
    \item \textbf{Feedback incorporation yields consistent gains} (+0.05 pass@k on average), primarily by reducing compilation errors.
    
    \item \textbf{In-context learning has negligible impact}, suggesting that the base prompt already provides sufficient structural guidance.
    
    \item \textbf{Model performance is language-dependent}: Python is the easiest, while Rust is consistently the most challenging due to stricter constraints.
    
    \item \textbf{Problem difficulty is the dominant factor}: performance drops sharply as complexity increases, with all models struggling on harder tasks.
\end{itemize}
\end{tcolorbox}


\subsection{Proximity to a Valid Solution}
\label{sec:results-proximity}

Table \ref{tab:proximity-metrics} reports the average CodeBLEU scores across languages and models, together with the individual component metrics. Several consistent trends emerge across languages and are later reflected in the final aggregated score. As mentioned before, similarity-based metrics were not computed for Rust as the tool used does not consistently support this language.

Python systematically obtains the lowest values across almost all components. This is expected: its flexible syntax and high-level constructs allow many different ways of expressing the same logic, which naturally reduces lexical and structural overlap with the reference solutions. Java typically comes next and exhibits consistently higher N-gram, weighted N-gram, and AST-matching scores. In contrast, C tends to score slightly lower on these components but achieves a higher dataflow-match score. This difference may be because these two languages use slightly different sets of problems due to the availability of correct reference solutions.

C++ generally reaches the highest proximity scores. This can be explained by the fact that the dataset contains a larger number of reference solutions in C++ compared to the other languages. Because we are computing the maximum similarity between a model output and all reference solutions for a given problem, having more references increases the chance of obtaining a closer match.

Some model-specific trends can also be observed. The larger proprietary models (Gemini-2.0-flash and GPT-4.1-mini) regularly achieve the highest scores across the multiple metrics, which aligns with their stronger correctness performance. Interestingly, when moving from plain N-gram to weighted N-gram matching, this contrast becomes even clearer, with the gap between the bigger proprietary models and the smaller open-source ones widening. Since weighted N-grams place more emphasis on informative tokens, this suggests that the stronger models more reliably reproduce the key lexical elements of the reference implementations.

The results also show that the model's relative performance is language-dependent. For instance, Qwen-2.5:14B is the best-performing open-source model in Python (0.49) yet becomes the weakest in Java (0.54), illustrating how a model’s relative strengths can shift substantially across languages.

Our results mirror the trends observed in the execution-based evaluations: stronger models generally produce solutions that are not only more correct but also more similar to the reference implementations. At the same time, proximity metrics provide additional insight into how models structure their solutions, beyond simple pass/fail outcomes. 
However, these trends should be interpreted with caution, given the known limitations of CodeBLEU. The metric favors solutions that are lexically and structurally similar to the reference implementations and may penalize functionally correct solutions that adopt alternative algorithms or coding styles. This is particularly evident in languages such as Python, where the same logic can be expressed in many different ways. Consequently, lower similarity scores do not necessarily imply lower solution quality, but may instead reflect diversity in valid implementations.

\begin{table}[t]
    \centering
    \caption{Proximity to a Valid Solution metrics}
    \label{tab:proximity-metrics}
    \begin{tabular}{ll|rrrr|r}
    \toprule
    Model & Lang. & Ngram & W. Ngram & AST & Dataflow & codeBleu \\
    \midrule
    \rowcolor{gray!15}
    deepseek-coder-v2 & Python & 0.10 & 0.19 & 0.56 & 0.55 & 0.49 \\
    \rowcolor{gray!15}
    gemini-2.0-flash & Python & 0.22 & 0.28 & 0.65 & 0.63 & 0.57 \\
    \rowcolor{gray!15}
    gpt-4.1-mini & Python & 0.18 & 0.28 & 0.67 & 0.64 & 0.58 \\
    \rowcolor{gray!15}
    qwen2.5-coder:14b & Python & 0.08 & 0.15 & 0.52 & 0.52 & 0.46 \\
    \rowcolor{gray!15}
    qwen2.5-coder:7b & Python & 0.07 & 0.14 & 0.51 & 0.49 & 0.43 \\
    \rowcolor{gray!15}
    qwen2.5:14b & Python & 0.14 & 0.21 & 0.57 & 0.54 & 0.49 \\ \hline
    deepseek-coder-v2 & C++ & 0.23 & 0.33 & 0.64 & 0.75 & 0.61 \\
    gemini-2.0-flash & C++ & 0.28 & 0.37 & 0.67 & 0.78 & 0.65 \\
    gpt-4.1-mini & C++ & 0.22 & 0.35 & 0.68 & 0.79 & 0.65 \\
    qwen2.5-coder:14b & C++ & 0.22 & 0.30 & 0.63 & 0.75 & 0.60 \\
    qwen2.5-coder:7b & C++ & 0.23 & 0.30 & 0.61 & 0.73 & 0.59 \\
    qwen2.5:14b & C++ & 0.20 & 0.28 & 0.62 & 0.75 & 0.60 \\ \hline
    \rowcolor{gray!15}
    deepseek-coder-v2 & Java & 0.25 & 0.35 & 0.68 & 0.61 & 0.57 \\
    \rowcolor{gray!15}
    gemini-2.0-flash & Java & 0.33 & 0.41 & 0.71 & 0.64 & 0.62 \\
    \rowcolor{gray!15}
    gpt-4.1-mini & Java & 0.27 & 0.42 & 0.71 & 0.64 & 0.61 \\
    \rowcolor{gray!15}
    qwen2.5-coder:14b & Java & 0.25 & 0.34 & 0.68 & 0.61 & 0.58 \\
    \rowcolor{gray!15}
    qwen2.5-coder:7b & Java & 0.24 & 0.33 & 0.67 & 0.59 & 0.56 \\
    \rowcolor{gray!15}
    qwen2.5:14b & Java & 0.23 & 0.31 & 0.66 & 0.56 & 0.54 \\ \hline
    deepseek-coder-v2 & C & 0.14 & 0.23 & 0.59 & 0.68 & 0.55 \\
    gemini-2.0-flash & C & 0.20 & 0.27 & 0.62 & 0.70 & 0.57 \\
    gpt-4.1-mini & C & 0.15 & 0.24 & 0.63 & 0.71 & 0.57 \\
    qwen2.5-coder:14b & C & 0.15 & 0.23 & 0.60 & 0.67 & 0.55 \\
    qwen2.5-coder:7b & C & 0.14 & 0.22 & 0.57 & 0.66 & 0.53 \\
    qwen2.5:14b & C & 0.14 & 0.21 & 0.58 & 0.68 & 0.54 \\
    \bottomrule
    \end{tabular}
\end{table}

\subsubsection{Feedback and ICL}

Feedback and \gls{ICL} have no significant impact on model performance under the proximity-based metrics. For \gls{ICL}, this is largely expected: we already observed that it does not meaningfully improve functional correctness, so it is unsurprising that the generated code remains highly similar.

The absence of improvement with feedback incorporation is also informative. Although models often adjust their outputs in response to the feedback, these edits tend to be small and localized (e.g., adding a missing import). Such minor corrections typically do not alter the broader syntactic structure or token distribution enough to affect CodeBLEU components. 

Because the observed differences are extremely small and exhibit no systematic trends, we do not report detailed tables for these conditions.

\subsubsection{Problem Difficulty}

Table~\ref{tab:codebleu-diff} shows how the values for CodeBleu and its components are affected by problem difficulty. Across all four CodeBLEU components, the scores tend to significantly decrease as problem difficulty increases. This pattern is visible in every language and for each metric. For example, the average N-gram score falls from 0.27 on the easiest problems to 0.07 on the hardest ones, and a similar drop appears in the weighted N-gram and AST components. The dataflow scores also decline, although the reduction is somewhat smaller for C++ and C, which have more uniform data-movement patterns across solutions. When looking at the final CodeBLEU score, the trend remains: easier tasks produce outputs that resemble the reference solutions more closely, both lexically and structurally. This suggests that, as tasks become more complex, models not only struggle to produce correct solutions but also generate solutions that are more distant from the human implementations.

\begin{table}[t]
\centering
\caption{CodeBleu values across difficulty levels}
\label{tab:codebleu-diff}

    \begin{subtable}[t]{0.1\textwidth}
        \setlength{\tabcolsep}{3.5pt}
        \centering
        \vspace{1.8em}
        \begin{tabular}{l}
        \toprule
        Lang. \\
        \midrule
        C \\
        C++ \\
        Java \\
        Python \\ \hline
        Avg. \\
        \bottomrule
        \end{tabular}
    \end{subtable}
    \hfill
    \begin{subtable}[t]{0.35\textwidth}
        \setlength{\tabcolsep}{3.5pt}
        \centering
        \caption{N-gram}
        \begin{tabular}{rrrr}
        \toprule
        0 & 1 & 2 & 3 \\
        \midrule
        0.22 & 0.09 & 0.05 & 0.05 \\
        0.32 & 0.18 & 0.12 & 0.09 \\
        0.36 & 0.20 & 0.12 & 0.08 \\
        0.19 & 0.09 & 0.06 & 0.04 \\ \hline
        0.27 & 0.14 & 0.09 & 0.07 \\
        \bottomrule
        \end{tabular}
    \end{subtable}
    \hfill
    \begin{subtable}[t]{0.35\textwidth}
        \setlength{\tabcolsep}{3.5pt}
        \centering
        \caption{Weighted N-gram}
        \begin{tabular}{rrrr}
        \toprule
        0 & 1 & 2 & 3 \\
        \midrule
        0.33 & 0.14 & 0.08 & 0.07 \\
        0.43 & 0.25 & 0.18 & 0.14 \\
        0.49 & 0.27 & 0.18 & 0.12 \\
        0.31 & 0.15 & 0.09 & 0.06 \\ \hline
        0.39 & 0.20 & 0.13 & 0.10 \\
        \bottomrule
        \end{tabular}
    \end{subtable}

    \vspace{1em}

    \begin{subtable}[t]{0.1\textwidth}
        \setlength{\tabcolsep}{3.5pt}
        \centering
        \vspace{1.8em}
        \begin{tabular}{l}
        \toprule
        Lang. \\
        \midrule
        C \\
        C++ \\
        Java \\
        Python \\ \hline
        Avg. \\
        \bottomrule
        \end{tabular}
    \end{subtable}
    \hfill
    \begin{subtable}[t]{0.27\textwidth}
        \setlength{\tabcolsep}{3.5pt}
        \centering
        \caption{AST}
        \begin{tabular}{rrrr}
        \toprule
        0 & 1 & 2 & 3 \\
        \midrule
        0.69 & 0.51 & 0.43 & 0.40 \\
        0.71 & 0.60 & 0.55 & 0.49 \\
        0.77 & 0.63 & 0.54 & 0.50 \\
        0.67 & 0.53 & 0.46 & 0.40 \\ \hline
        0.71 & 0.57 & 0.50 & 0.45 \\
        \bottomrule
        \end{tabular}
    \end{subtable}
    \hfill
    \begin{subtable}[t]{0.27\textwidth}
        \setlength{\tabcolsep}{3.5pt}
        \centering
        \caption{Dataflow}
        \begin{tabular}{rrrr}
        \toprule
        0 & 1 & 2 & 3 \\
        \midrule
        0.83 & 0.55 & 0.42 & 0.36 \\
        0.89 & 0.70 & 0.58 & 0.45 \\
        0.80 & 0.49 & 0.34 & 0.24 \\
        0.70 & 0.49 & 0.37 & 0.27 \\ \hline
        0.80 & 0.56 & 0.43 & 0.33 \\
        \bottomrule
        \end{tabular}
    \end{subtable}
    \hfill
    \begin{subtable}[t]{0.27\textwidth}
        \setlength{\tabcolsep}{3.5pt}
        \centering
        \caption{CodeBleu}
        \begin{tabular}{rrrr}
        \toprule
        0 & 1 & 2 & 3 \\
        \midrule
        0.66 & 0.45 & 0.35 & 0.32 \\
        0.71 & 0.56 & 0.48 & 0.40 \\
        0.71 & 0.50 & 0.38 & 0.32 \\
        0.61 & 0.44 & 0.35 & 0.28 \\ \hline
        0.68 & 0.49 & 0.39 & 0.33 \\
        \bottomrule
        \end{tabular}
    \end{subtable}

\end{table}

\begin{tcolorbox}[colback=gray!10, colframe=black, title=\textbf{Key Takeaways}]
\begin{itemize}
    \item \textbf{Stronger models produce solutions closer to reference implementations}, mirroring trends observed in functional correctness.
    
    \item \textbf{Language characteristics influence similarity scores}: Python exhibits lower similarity due to higher implementation diversity, while C++ achieves the highest scores.
    
    \item \textbf{Feedback and ICL have minimal impact}, indicating that corrections are mostly local and do not alter overall structure.
    
    \item \textbf{Similarity decreases with problem difficulty}, suggesting that harder tasks lead to more divergent and less structured solutions.
\end{itemize}
\end{tcolorbox}

\subsection{Code Quality}
\label{sec:results-quality}

\subsubsection{Baseline}

\begin{figure}
    \centering
    \includegraphics[width=1.0\linewidth]{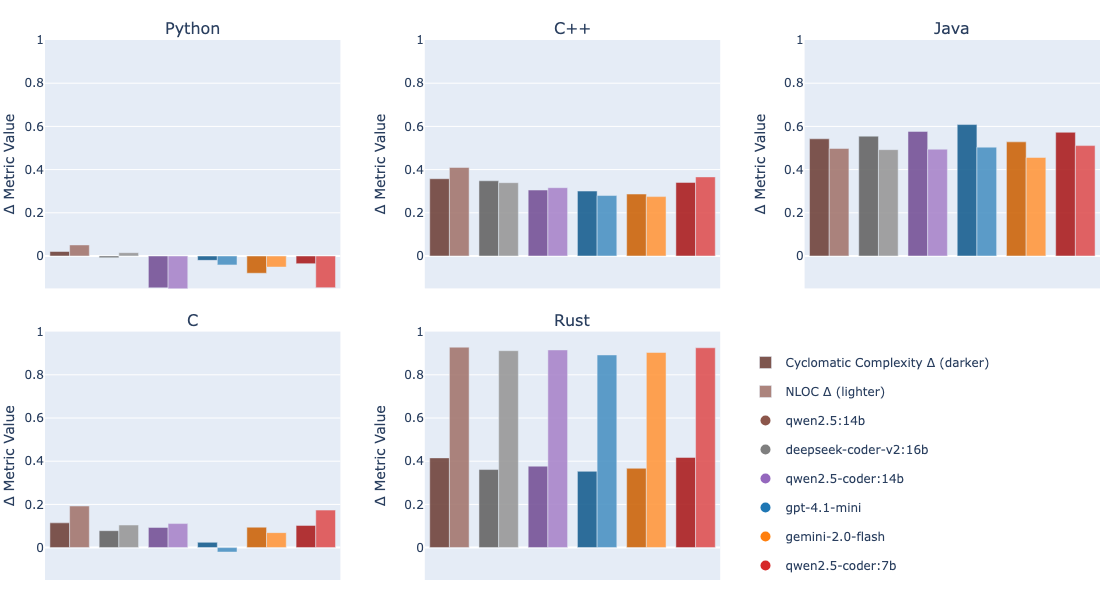}
    \caption{Code Quality Metrics}
    \label{fig:code_quality}
\end{figure}

Figure \ref{fig:code_quality} shows that the code quality metrics reveal clear differences across both programming languages and model families. Python is the language in which LLM-generated solutions most closely resemble the reference implementations. Qwen2.5:14b produces slightly simpler Python solutions while the others generate solutions that are either as complex or more than the reference.

This pattern does not generalize to the remaining languages. In Rust, every model produces solutions that are substantially simpler according to both metrics, with uniformly large positive values. This tells us that LLM-generated Rust code tends to be much less verbose and structurally simpler than the reference implementations. C++ and Java exhibit a similar pattern, though with smaller magnitudes, while C shows moderate positive deviations across all models except GPT-4.1-mini, whose scores remain nearly neutral.

We believe these differences are not solely attributable to the models but also reflect characteristics of the underlying dataset. For example, reference Python solutions may have been written more carefully or with more deliberate structure, potentially because Python submissions run more slowly, making users be more careful when writing the code.

Model-specific patterns also emerge. GPT-4.1-mini, which achieves the lowest values for both metrics in C, shows the highest values in Java. This suggests that its code-generation behavior depends on the specific conventions and idioms of each language.

Finally, although cyclomatic complexity and \gls{NLOC} generally correlate, examining both metrics provides additional insights. For instance, focusing on the C++ results, GPT-4.1-mini shows relatively higher increases in cyclomatic complexity than in \gls{NLOC}, whereas Qwen2.5:14b shows the opposite pattern. This indicates that GPT-4.1-mini tends to introduce more branching structure without proportionally increasing the amount of code, while Qwen2.5:14b produces more verbose code.

Overall, these results show the importance of including code quality metrics in LLM benchmarks, as they reveal differences in model behavior that are invisible to traditional correctness-based evaluations.

\subsubsection{ICL and Feedback-incorporation}

Regarding the effect of \gls{ICL} and feedback on the quality of the generated code, the results showed that, across all models and languages, the differences relative to the baseline prompting strategy were negligible, both for cyclomatic complexity and nloc. These results are consistent with the earlier findings on functional correctness: since ICL did not improve pass@k performance, it is unsurprising that it also fails to alter the structural characteristics of the solutions. For feedback, the outcome is similarly expected. The feedback mechanism used in our evaluation provides information about functional validity rather than stylistic or structural quality, so it does not meaningfully affect how models choose to structure their implementations once they produce a correct solution. 

\subsubsection{Problem Difficulty}

By looking at the results in Table \ref{tab:quality-diff}, we see that the generated solutions for the hardest problems tend to have much lower cyclomatic complexity and fewer lines of code than the corresponding human-written reference implementations. At first glance this could suggest that the models produce code that is extremely simple for the most difficult tasks. However, this pattern is not very reliable because most models only managed to solve a handful of the hardest problems. With such small sample sizes, the averages for Difficulty 3 have high variance and should be interpreted with caution.

A clearer picture comes from the model that solved the largest number of high-difficulty tasks, which in our case is gpt-4.1-mini. Even though the number of solved problems in this category is still limited, its complexity and NLOC values are already closer to those observed for the easier difficulty levels. This indicates that the apparent simplicity gap for difficult problems is mostly a consequence of the small number of successful solutions rather than a consistent model behavior.

Overall, these results point to a pattern that is similar to what we observed for ICL and Feedback. Problem difficulty does not seem to have a strong or systematic influence on the quality of the generated code when compared with human-written reference implementations. Any deviations at the highest difficulty level appear to stem from insufficient data rather than from meaningful differences in code quality.

\begin{table}[t]

\centering
\caption{Code quality metrics across difficulty levels}
\label{tab:quality-diff}

    \begin{subtable}[t]{0.25\textwidth}
        \centering
        \vspace{1.8em}
        \begin{tabular}{l}
        \toprule
        Diff. Level \\
        \midrule
        deepseek-coder-v2 \\
        gemini-2.0-flash \\
        gpt-4.1-mini \\
        qwen2.5-coder:14b \\
        qwen2.5-coder:7b \\
        qwen2.5:14b \\ \hline
        Average \\
        \bottomrule
        \end{tabular}
    \end{subtable}
    \hfill
    \begin{subtable}[t]{0.35\textwidth}
        \centering
        \caption{Cyclomatic complexity}
        \begin{tabular}{rrrr}
        \toprule
        0 & 1 & 2 & 3 \\
        \midrule
        0.21 & 0.22 & 0.29 & 0.80 \\
        0.25 & 0.17 & 0.19 & 0.60 \\
        0.28 & 0.19 & 0.19 & 0.38 \\
        0.19 & 0.18 & 0.22 & 0.70 \\
        0.21 & 0.25 & 0.15 & 0.74 \\
        0.27 & 0.25 & 0.25 & 0.75 \\ \hline
        0.24 & 0.21 & 0.22 & 0.66 \\  
        \bottomrule
        \end{tabular}
    \end{subtable}
    \hfill
    \begin{subtable}[t]{0.35\textwidth}
        \centering
        \caption{NLOC}
        \begin{tabular}{rrrr}
        \toprule
        0 & 1 & 2 & 3 \\
        \midrule
        0.31 & 0.36 & 0.44 & 0.78 \\
        0.31 & 0.29 & 0.31 & 0.60 \\
        0.33 & 0.28 & 0.26 & 0.41 \\
        0.26 & 0.29 & 0.34 & 0.68 \\
        0.29 & 0.35 & 0.40 & 0.74 \\
        0.38 & 0.41 & 0.45 & 0.75 \\ \hline
        0.31 & 0.33 & 0.37 & 0.66 \\
        \bottomrule
        \end{tabular}
    \end{subtable}

\end{table}

\begin{tcolorbox}[colback=gray!10, colframe=black, title=\textbf{Key Takeaways -- Code Quality}]
\begin{itemize}
    \item \textbf{LLM-generated code is generally simpler and shorter} than the human-written solutions, except in Python.
    
    \item \textbf{Quality patterns are language-dependent}, with notably simpler outputs in Rust, C++, and Java.
    
    \item \textbf{Model behavior varies by language}: some models favor more branching logic, while others produce more verbose code.
    
    \item \textbf{ICL and feedback do not affect code structure}, reinforcing that these techniques target correctness rather than quality.
    
    \item \textbf{Difficulty has no clear systematic effect on quality}, with apparent trends at high difficulty largely driven by small sample sizes.
\end{itemize}
\end{tcolorbox}

\subsection{Common Problems in the Generated Code}
\label{sec:results-common-problems}

To complement the quantitative metrics, we also conducted a manual review of the generated code to better understand the recurring issues behind the failures. For this analysis, we examined four possible unit-test failure outcomes: compilation error, runtime error, timeout, and incorrect output. 

For each category, we randomly sampled multiple generated solutions across different models and programming languages and manually inspected them to identify recurring patterns and representative failure modes. In total, we analyzed approximately 30 samples per failure category, which, while not statistically exhaustive, was sufficient to reveal consistent and recurring patterns.

For each category, we begin with an overview of the most common causes and then provide a few illustrative examples we observed.

\subsubsection{Compilation Error}

Compilation errors are strongly language-dependent, but several recurring patterns appeared across models. Some issues are universal, such as calling functions with the wrong parameter types or referencing variables that had never been declared (or had been declared only inside a narrower scope). In compiled languages such as C, C++, and Rust, another common source of failure was using functions or types without including the appropriate headers or modules. These mistakes show that the models can still struggle with very basic programming fundamentals.

Two different compilation error examples can be seen in Listings~\ref{lst:ex-comp-1} and~\ref{lst:ex-comp-2}. Listing~\ref{lst:ex-comp-1} shows a C++ snippet in which \texttt{std::function} is used without including the corresponding header \texttt{<functional>}, causing an immediate compilation error. Listing~\ref{lst:ex-comp-2} illustrates a Java example where the method \texttt{modInverse()} attempts to return \texttt{x \% MOD}, even though \texttt{MOD} is not defined within its scope (\texttt{MOD} was declared only inside \texttt{calculateProduct()}).

\begin{figure}
\begin{lstlisting}[style=smallcpp, caption={Compilation error caused by a missing import}, label={lst:ex-comp-1}]
#include <iostream>
#include <vector>
#include <algorithm>
using namespace std;

int main() {
    ...
    function<void(int, int)> dfs = [&](int node, int parent) {
        ...
        }
...
\end{lstlisting}
\end{figure}

\begin{figure}
\begin{lstlisting}[style=smalljava, caption={Compilation error caused by attempting to access an undefined variable}, label={lst:ex-comp-2}]
...
public class ArithmeticProgressionProduct {
    ...
    private static long calculateProduct(long x, long d, long n) {
        final int MOD = 1000003;
        ...
    }
    
    private static long modInverse(long a, int m) {
        ...
        return x % MOD;
    }
\end{lstlisting}    
\end{figure}

\subsubsection{Runtime Error}

A similar picture emerged from the runtime errors. In Python, many failures were due to missing imports or incorrect variable scoping, issues that would have been caught at compile time in the other languages. Errors such as out-of-bounds accesses, incorrect input handling, or excessive recursion appeared across all languages. In C and C++, additional failures frequently were caused by attempting to allocate extremely large arrays on the stack or heap, which caused segmentation faults or memory-allocation failures at runtime. Rust stood out for producing a higher number of overflow-related crashes: because Python uses arbitrary-precision integers and C/C++/Java typically allow overflow to occur silently and merely propagate incorrect values, only Rust, whose standard integer types panic on overflow in debug mode, registered these cases as runtime failures rather than silent wrong answers.

Some of these problems can be seen in the examples shown in the Listings~\ref{lst:ex-runtime-1}, \ref{lst:ex-runtime-2}, \ref{lst:ex-runtime-3}. In the first example (Listing~\ref{lst:ex-runtime-1}), the code generated by the \gls{LLM} attempted to create an array of size \(N(N-1)/2\). For one unit test, the value of N was 100,000, meaning that the program tried to allocate on the order of \(5\times 10^{9}\) entries, which led to a segmentation fault due to failed allocation (required too much memory).
The second example (Listing~\ref{lst:ex-runtime-2}) shows a program that fails due to an off-by-one indexing error. The table \texttt{m} is allocated as an \(n \times n\) matrix, so valid column indices range from \(0\) to \(n-1\). However, the loop eventually makes \texttt{j = n} and then attempts to write to \texttt{m[i][j]}, an index just past the end of the array. 
The final example (Listing~\ref{lst:ex-runtime-3}) displays a Rust code sample that crashed due to integer overflow while computing \texttt{d.pow(n)}.

\begin{figure}[h]
\begin{lstlisting}[style=smallcpp, caption={Runtime error caused by invalid memory allocation}, label={lst:ex-runtime-1}]
...
int main() {
    int N, M;
    cin >> N >> M;
    vector<bool> even_edges(N * (N - 1) / 2, false);
...
\end{lstlisting}
\end{figure}

\begin{figure}
\begin{lstlisting}[style=smallrust, caption={Runtime error caused by out-of-bounds access to an array}, label={lst:ex-runtime-2}]
...
let mut m = vec![vec![0; n]; n];
for l in 2..n {
    for i in 0..=n-l {
        let j = i + l;
        m[i][j] = usize::MAX;
        ...
\end{lstlisting}
\end{figure}

\begin{figure}[h]
\begin{lstlisting}[style = smallrust, caption={Runtime error caused by an Overflow}, label={lst:ex-runtime-3}]
fn main() {
    ...
    let n = parts[0]; // parts is read from the input
    ...
    for d in (1..=p).rev() {                     
        if p % d == 0 && d.pow(n as u32) <= p {
            ...
        }
    }
    println!("{}", max_gcd);
}
\end{lstlisting}
\end{figure}

\subsubsection{Timeout}

Timeout errors were often caused by inefficient algorithms and logical issues. Models sometimes defaulted to brute-force enumeration or repeatedly performed expensive operations such as full-array reversals, sorts, or data-structure rebuilds after every update. 

In the example shown in Listing~\ref{lst:ex-timeout-1}, the model was asked to provide code to count the number of triples \((a,b,c)\) with \(1\le a,b,c\le N\) such that \(a+b\), \(b+c\), and \(c+a\) are all multiples of \(K\). The \gls{LLM}-generated solution attempted to check all \(N^3\) possible triples. However, all three conditions together imply that \(a\), \(b\), and \(c\) must share the same remainder when divided by \(K\), meaning that the answer can be computed simply by counting how many values fall into each remainder class and then taking combinations, yielding a much more efficient solution.

In the example shown in Listing~\ref{lst:ex-timeout-2}, the task was to construct a sequence produced by repeatedly appending elements and reversing the entire list after each append. The \gls{LLM}-generated solution reversed the whole list every time, resulting in an \(O(n^2)\) algorithm. The correct approach is to observe that each operation merely flips the orientation of the sequence: rather than performing full reversals, one tracks whether the sequence is currently “forward” or “reversed” and inserts new elements at the appropriate end. Printing the resulting sequence in the correct direction at the end yields the right answer in \(O(n)\) time.

\begin{figure}[h]
\begin{lstlisting}[style=smallcpp, caption={Timeout error caused by using an extremelly inefficient approach}, label={lst:ex-timeout-1}]
...
for (int a = 1; a <= N; a++) {
    for (int b = 1; b <= N; b++) {
        for (int c = 1; c <= N; c++) {
            if ((a + b) % K == 0 && (b + c) % K == 0 && (c + a) % K == 0) {
                count++;
            }
...
\end{lstlisting}
\end{figure}

\begin{figure}[h]
\begin{lstlisting}[style=smallcpp, caption={Timeout Error caused by repeatedly performing a costly operation}, label={lst:ex-timeout-2}]
...
for (int i = 0; i < n; i++) {
    // Append a_i to the end of b
    b[b_size++] = a[i];
    
    // Reverse the order of the elements in b
    for (int j = 0; j < b_size / 2; j++) {
        int temp = b[j];
        b[j] = b[b_size - j - 1];
        b[b_size - j - 1] = temp;
    }
}
\end{lstlisting}
\end{figure}

\subsubsection{Incorrect Output}

For incorrect outputs, the issues were more varied and not concentrated in just a few simple patterns, so there were not as many clearly shared root causes. In general, many errors were due to the model misreading or oversimplifying the problem’s structure (e.g., ignoring key constraints), wrong overall logic, and failing to consider some edge cases. 

In the first example (Listing~\ref{lst:ex-incorrect-1}), the model was asked to simulate an alternating draft where two players choose dishes to maximize the difference between their total happiness (each dish gives different happiness to each player). This means that the importance of each dish depends not only on what the chooser gains but also on what the opponent would gain if allowed to take it. The correct priority of a dish is therefore based on $A_i + B_i$, since choosing it both earns the chooser $A_i$ (or $B_i$) and prevents the opponent from earning $B_i$ (or $A_i$). However, the model incorrectly sorted dishes by $A_i - B_i$, which measures only the direct scoring difference and ignores the denial effect, causing it to select dishes in the wrong order and thus an incorrect final outcome.

In the second example (Listing~\ref{lst:ex-incorrect-2}), the models are asked to determine whether \(\sqrt{a} + \sqrt{b} < \sqrt{c}\). The model solved this by directly computing the square roots using floating-point arithmetic and comparing 
the results. This approach usually works for ordinary inputs, but, for one unit test, the values were extremely large and the two sides of the inequality almost equal. Because floating-point numbers cannot represent such tiny differences accurately, the comparison becomes unreliable. As a result, the model outputs the wrong answer.

\begin{figure}[h]
\begin{lstlisting}[style=smallpy, caption={Incorrect output caused by flawed logic}, label={lst:ex-incorrect-1}]
...
# Sort the dishes by the difference of happiness points (A-B).
sorted_dishes = sorted(AB_pairs, key=lambda x: x[0] - x[1], reverse=True)
...
\end{lstlisting}
\end{figure}

\begin{figure}[h]
\begin{lstlisting}[style=smallcpp, caption={Incorrect output due to missing edge unit test}, label={lst:ex-incorrect-2}]
int main() {
    long double a, b, c;
    scanf("%Lf %Lf %Lf", &a, &b, &c);
    
    if(sqrt(a) + sqrt(b) < sqrt(c)) 
        printf("Yes\n");
    else
        printf("No\n");

    return 0;
}
\end{lstlisting}
\end{figure}

\section{Discussion and Threats to Validity}
\label{sec:dicusiion-ttv}

\subsection{Discussion}
\label{sec:discussion}

Our analysis reveals several consistent trends about the current capabilities and limitations of state-of-the-art \glspl{LLM} in text-to-code generation.

\textbf{Larger models consistently achieve higher functional correctness, but even the strongest systems remain far from reliable.} Although the bigger, proprietary models consistently achieved higher scores for the functional correctness metrics than the smaller, open-source models, these models still fail frequently, especially in harder tasks, showing that scale alone is not enough.

\textbf{Feedback incorporation substantially improves correctness, while \gls{ICL} has negligible effect.}
Compiler feedback allows models to correct simple, localized issues (e.g., missing imports, type mismatches), but in-context examples add little because the base prompt already provides detailed structural instructions.

\textbf{All models struggle sharply as problem difficulty increases.} Performance drops across all metrics, and errors become more frequent, indicating that current \glspl{LLM} have limited ability to generalize their reasoning to more complex problems.

\textbf{Models that produce more correct code also tend to be lexically and structurally closer to reference solutions.} The proximity-based metrics show a clear alignment between correctness and similarity, suggesting that good solutions not only execute correctly but also mirror human implementations more closely.

\textbf{Proximity to reference solutions is largely unaffected by feedback.} Although the functional correctness metrics all clearly improved with feedback, the same cannot be said for the proximity metrics. This shows that no big structural changes to the code were made, meaning the corrections applied after receiving compiler messages are small and localized, leaving the overall structure and high-level organization of the generated solutions essentially unchanged.

\textbf{Generated code is generally less complex and shorter than human-written solutions, though the effect is language-dependent.} When looking at the code-quality metrics, we can see that the generated code is simpler than the reference solution in all languages except Python, showing the \gls{LLM}-generated code tends to be simpler.

\textbf{Generated code is often unreliable, exhibiting fundamental and easily avoidable errors.}
Models frequently produce code with severe issues, such as allocating data structures without verifying that the required memory is feasible, performing operations that trigger integer overflows, accessing arrays out of bounds, or relying on brute-force algorithms that exceed time or memory limits. These failures highlight the risks of using \glspl{LLM} for code generation, as developers may overtrust automatically generated solutions, potentially allowing dangerous vulnerabilities or performance regressions to reach production systems.

Overall, these findings show that current \glspl{LLM} remain far from reliably solving diverse and complex programming tasks, even when provided with detailed prompts and corrective feedback. The models still make very basic mistakes, generate code that can be unreliable or unsafe, and frequently fail to handle edge cases or problem-specific constraints. These limitations reinforce the need for improved reasoning capabilities, stronger error-detection and self-repair mechanisms, and training data that better captures the syntactic, semantic, and algorithmic diversity of real-world programming tasks. They also highlight the practical value of our benchmark in providing a more comprehensive assessment of \glspl{LLM}’ coding abilities and supporting future research in this area.

\subsection{Threats to Validity}
\label{sec:ttv}

\textbf{Internal Validity}: One potential threat is that the evaluated \glspl{LLM} may have encountered some of the benchmark problems during pretraining, which could artificially inflate performance. Our dataset originates from 2021, and although it is possible that some problems were seen by the models, their overall limited performance suggests that any such memorization had a negligible impact on the results.

\textbf{External Validity}: There is a wide variety of \glspl{LLM} with different sizes, architectures, training objectives, and intended applications. As such, generalizing our findings to models not included in this study is uncertain. To reduce this risk, we selected a diverse set of models, ranging from small open-source to large closed-source ones, covering both code-specific and general-purpose models, as well as dense and \gls{MoE} architectures. 

A further limitation is that all benchmark problems were self-contained competitive programming problems, which may not fully reflect real-world coding scenarios that involve larger codebases or interdependent components. However, the problems included diverse input types and were sufficiently challenging, with models failing to achieve consistently high performance. This indicates the tasks were non-trivial and, by spanning varied formats and complexities, still provide meaningful insights despite being self-contained. In addition, the PROBE framework is designed to be extensible to more realistic software development settings. Following the guidelines in Section~\ref{sec:extensibility}, the benchmark can be extended to tasks involving larger codebases or multi-file projects by incorporating new task descriptions, providing corresponding unit tests, and, when required, including reference implementations to enable proximity and quality evaluation. This would allow future work to assess scenarios such as code completion, bug fixing, or repository-level reasoning, while preserving the same evaluation principles.

\textbf{Construct Validity}: Evaluating similarity to reference solutions is inherently limited. While multiple reference solutions are available for most problems in each programming language, it is impossible to cover every valid implementation. Consequently, a generated solution may appear distant from the references simply because the corresponding implementation is absent from the dataset. To reduce this issue, we compute similarity metrics only for problems with at least three reference solutions per language, though the limitation remains. Another limitation concerns code quality assessment: although we consider multiple aspects using two different metrics, other dimensions, such as maintainability and readability, are not fully captured. 

A further threat arises from the use of LLM-generated unit tests. Although synthetic inputs enable validation in cases where no test cases are available in the original dataset, they may introduce bias or fail to fully capture the problem specification, particularly for edge cases. To mitigate this, we use LLMs only to generate test inputs, while expected outputs are derived by executing multiple reference implementations, retaining only those test cases for which all implementations agree. In addition, we assess test adequacy using coverage metrics to ensure that the resulting test suites exercise a substantial portion of the program logic. While these measures reduce the risk of incorrect or misleading evaluation signals, they do not fully eliminate the possibility of missing corner cases.

Finally, problem difficulty is estimated based on static code metrics derived from reference implementations, which reflect implementation complexity rather than intrinsic problem difficulty, and may therefore be influenced by coding style or algorithmic choices. Although we mitigate this by averaging across multiple solutions and observe consistent performance degradation as difficulty increases, the resulting categorization should be interpreted as a proxy rather than a ground-truth measure.

\section{Conclusion}
\label{sec:conclusion}

In this paper, we introduced \texttt{PROBE}, a framework for benchmarking \glspl{LLM} on text-to-code generation. The framework builds on a diverse dataset of 1,651 problems spanning five programming languages and covering a range of difficulty levels. Generated code is assessed along three complementary dimensions: functional correctness, proximity to a valid solution, and code quality. To demonstrate its use, we applied \texttt{PROBE} to six models, including both open-source and proprietary systems, as well as general-purpose and code-specialized variants.

Our evaluation showed that performance strongly depends on programming language and model size, with Python, C++, and Java generally yielding better results, while smaller models struggle significantly in lower-resource languages such as Rust. We further showed that feedback-based strategies consistently improve performance, whereas in-context learning provides only marginal benefits. Finally, we observed that as problem difficulty increases, correctness and proximity to valid solutions decline, but the gap in code quality between human-written and model-generated does not change significantly. In addition, our error analysis revealed that models frequently fail due to fundamental and easily avoidable issues, such as out-of-bounds accesses, integer overflows, and inefficient brute-force solutions, underscoring that even seemingly correct outputs may be unreliable in practice. These findings highlight the need for more robust and nuanced benchmarks to guide future progress in text-to-code generation.

As future work, we plan to extend the framework by covering additional programming languages and continuously incorporating newly released state-of-the-art models across different providers, ensuring that the benchmark remains up-to-date as the field evolves. This includes, but is not limited to, models from families such as LLaMA and Claude. We also aim to include new types of problems that require integration with existing code, enabling a more comprehensive assessment of models in realistic development scenarios. Finally, we plan to expand the set of code quality metrics by incorporating maintainability and readability-related metrics, allowing for a more comprehensive assessment of the generated code.

\section*{Declarations}
\addcontentsline{toc}{section}{Declarations}



\subsection*{Ethical approval}

Not applicable.

\subsection*{Informed consent}

Not applicable.

\subsection*{Author Contributions}

All authors (Rodrigo Pato Nogueira, Marco Vieira, and João R. Campos) contributed equally to this paper.

\subsection*{Data Availability Statement}

The complete dataset, including problem descriptions, unit tests, and reference solutions, as well as all the code used for both the code generation and evaluation are publicly available at \url{https://huggingface.co/datasets/OSS-forge/PROBE}.

\subsection*{Conflict of Interest }

The authors declare that they have no conflict of interest.



\bibliographystyle{spbasic}      
\bibliography{bibliography}   

\end{document}